%% file: H1B_MortgageLockin_5_20_2026.tex
\documentclass[12pt]{article}

\usepackage[margin=1in]{geometry}
\usepackage{setspace}
\usepackage{palatino}
\usepackage[utf8]{inputenc}
\usepackage[T1]{fontenc}
\usepackage{pdflscape}  
\usepackage{adjustbox}
\usepackage[hang,flushmargin]{footmisc}
\usepackage{amsmath}
\usepackage{amssymb}
\usepackage{amsthm}

\usepackage{booktabs}
\usepackage{graphicx}
\usepackage{caption}
\usepackage{subcaption}
\usepackage{rotating}
\usepackage{multirow}
\usepackage{threeparttable}
\usepackage[labelfont=bf, labelsep=newline, justification=centering, singlelinecheck=false]{caption}

\usepackage{natbib}
\usepackage{hyperref}
\hypersetup{
    colorlinks=true,
    linkcolor=blue,
    citecolor=blue,
    urlcolor=blue
}

\usepackage{fancyhdr}
\usepackage{titlesec}
\titleformat{\section}{\normalfont\large\bfseries}{\thesection.}{1em}{}
\titleformat{\subsection}{\normalfont\normalsize\bfseries}{\thesubsection}{1em}{}

\begin{document}

\begin{titlepage}
\begin{center}
\vspace*{1cm}

{\LARGE\bfseries Mobile Foreigners: \\ Mortgage Lock-In and H-1B Demand}

\vspace{0.8cm}

\renewcommand{\thefootnote}{\fnsymbol{footnote}}
{\large Duha T. Altindag\footnotemark[1] \hspace*{.15cm} John M. Nunley\footnotemark[2] \hspace*{.15cm} and \hspace*{.15cm} R. Alan Seals\footnotemark[3]}

\footnotetext[1]{Auburn University, \href{mailto:altindag@auburn.edu}{altindag@auburn.edu}}
\footnotetext[2]{University of Wisconsin--La Crosse, \href{mailto:jnunley@uwlax.edu}{jnunley@uwlax.edu}}
\footnotetext[3]{Auburn University, \href{mailto:alan.seals@auburn.edu}{alan.seals@auburn.edu}}

\vspace{0.3cm}

{\normalsize \today}

\vspace{0.8cm}

{\normalsize\bfseries Abstract}
\end{center}

\begin{quote}

\noindent The 2022 rise in U.S.\ mortgage rates increased relocation costs for homeowners with low-rate mortgages. This cost varies across destinations because each draws workers from a different mix of labor markets. We build an in-migration mortgage-payment wedge from HMDA loans and pre-shock IRS migration networks. From 2017 to 2024, higher wedges reduce college-educated homeowner in-migration, leave renters unaffected, and raise H-1B sponsorship requests. The implied offset is 14 H-1B sponsorship requests per 100 deterred college-educated domestic in-migrants. We show that mortgage lock-in operates as a destination-side labor-market shock that shifts part of firms' adjustment toward employer-sponsored immigration.
\end{quote}

\vspace{0.5cm}


\vspace{0.3cm}


\vfill


\end{titlepage}
\thispagestyle{empty}  

\begin{center}
{\normalsize\bfseries Extended Abstract}
\end{center}

\noindent The 2022 rise in U.S.\ mortgage rates increased the cost of moving for homeowners holding low-rate mortgages. We study how this national shock was transmitted unevenly across local high-skilled labor markets through destinations' historical migration networks. Some destinations draw movers from origins where homeowners retain low-rate mortgages with large payment advantages, while others draw from origins with smaller locked-in advantages. We capture this variation with a destination-level in-migration mortgage-payment wedge, $MPW^{in}$, built from HMDA mortgage-originations data, a 2024 mortgage-pricing rule applied to 2020 and 2021 borrowers, and pre-shock IRS migration flows.

In a 2017 to 2024 commuting-zone panel, higher $MPW^{in}$ reduces domestic high-skilled in-migration after the rate shock. The decline is concentrated among college-educated homeowners, while college-educated renters show no response, supporting a mortgage-lock-in interpretation rather than a general decline in destination attractiveness. In the same destinations, realized inflows of college-educated migrants from abroad rise, and employers expand H-1B sponsorship, with newly requested workers rising about 20 percent of the pre-period mean. USCIS petition and approval outcomes also rise, although by less. Event studies place the divergence after 2022. An instrumental-variables strategy isolating the feeder-origin component of $MPW^{in}$ reproduces the estimates with a strong first stage, and placebo tests support a causal reading.

The H-1B response is larger where the wedge coincides with occupation-specific labor demand, with a positive triple interaction between $MPW^{in}$, the post-2022 period, and predicted SOC demand growth. Decomposing $MPW^{in}$ shows the migration response is driven primarily by feeder-origin lock-in, while the H-1B response reflects both destination and origin payment conditions. We estimate roughly 14 additional H-1B sponsorship requests and 3 approved workers per 100 deterred college-educated domestic in-migrants, indicating that mortgage lock-in shifts part of firms' adjustment toward employment-based immigration.

\clearpage
\setcounter{page}{1}

\section{Introduction}\label{sec:intro}

The effects of U.S.\ monetary policy are regionally heterogeneous \citep{carlino_differential_1998, beraja_regional_2019}, and the local channels driving that variation are hard to identify \citep{romer_new_2004, nakamura_identification_2018}. Monetary tightening pushed the average 30-year fixed mortgage rate from a record low of 2.65 percent in January 2021 to 7.79 percent by October 2023, the highest level in more than two decades.\footnote{Weekly average 30-year fixed mortgage rate from the Freddie Mac Primary Mortgage Market Survey, accessed via FRED series \texttt{MORTGAGE30US}, \url{https://fred.stlouisfed.org/series/MORTGAGE30US}.} Although the interest rate shock was nationwide, local labor markets differ in how many of their homeowners hold low-rate fixed mortgages, and those homeowners face a substantial cost gap to move under the higher rate regime. Crucially, that gap varies across destinations, because each destination attracts workers from a different mix of labor markets. Using this cross-destination variation, we show that mortgage lock-in constrains domestic high-skilled mobility and shifts part of firms’ labor-market adjustment toward employer-sponsored immigration.

Existing empirical studies of mortgage lock-in typically focus on household mobility and ask how much homeowners give up by moving \citep{ferreira_housing_2010,ioannides_housing_2025, howard_jue_2026, fonseca_unlocking_2024, howard_effects_2025}. We instead ask whether a destination labor market can attract the domestic workers it normally draws through internal migration. Our measure, the in-migration mortgage-payment wedge ($MPW^{in}$), captures the expected increase in monthly mortgage payments a worker would face when moving into a destination, averaged across the origin locations that historically supply migrants to that destination. We construct $MPW^{in}$ by weighting HMDA-based mortgage-payment wedges with pre-shock IRS county-to-county migration flows.

Variation in $MPW^{in}$ across destinations comes from the composition of their feeder origins, not from local housing costs at the destination alone. For example, a high-payment destination can have a low wedge if its feeder origins are also high-payment places. Similarly, a moderately priced destination may have a high $MPW^{in}$ if its feeder origins are populated by homeowners holding low-rate mortgages with large payment advantages. Thus, although the post-2021 increase in mortgage rates was a national phenomenon, each destination's in-migration costs vary with its migration network. In this sense, $MPW^{in}$ is a shift-share measure of exposure to moving costs, with the post-2021 national increase in mortgage rates as the shock and destination-specific feeder-origin networks as the shares.

 A higher $MPW^{in}$ makes domestic homeowners in origin markets less responsive to job opportunities at potential destinations. The friction therefore matters most when firms in a destination have positions to fill, because higher moving costs make it harder to draw domestic high-skilled workers from other local labor markets.\footnote{About 65 percent of U.S.\ households are owner-occupied (\url{https://fred.stlouisfed.org/series/RSAHORUSQ156S}), so mortgage-payment frictions shape relocation decisions for a large share of the workforce. Worker migration has long been the primary channel through which local labor markets adjust to demand shocks \citep{blanchard_regional_1992}, and the secular weakening of this channel has been linked to origin-side housing constraints \citep{olney_determinants_2024}.} Firms may then adjust through wages, recruiting intensity, remote work, delayed hiring, or employment-based immigration sponsorship. \citet{cadena_immigrants_2016} show that when one group of workers is geographically immobile, more mobile foreign-born workers can absorb local shocks and reduce the incidence on natives. We study one observable firm-side margin of this adjustment: whether employers increase H-1B sponsorship in destinations where mortgage lock-in makes domestic high-skilled in-migration less responsive.

Our commuting-zone (CZ) panel covers 2017 to 2024 and combines ACS migration tabulations with H-1B Labor Condition Applications (LCAs) that measure employer sponsorship demand and downstream USCIS petition and approval outcomes. High-wedge CZs receive fewer college-educated migrants after 2022, and the decline is concentrated among college-educated homeowners. College-educated renters, who are not exposed to mortgage lock-in, show no response. High-wedge destinations also receive more college-educated inflows from abroad, and this increase is concentrated among non-citizens rather than returning citizens. In the same high-wedge CZs, H-1B sponsorship activity rises, with firms requesting more H-1B workers. More distinct firms file new-employment LCAs in high-wedge destinations. This firm-count response indicates that the increase is not just a larger-request margin among incumbent sponsors. Downstream USCIS adjudication data corroborate the pattern, as H-1B visa approvals also increase in high-wedge CZs after 2022. A one-standard-deviation increase in $MPW^{in}$ lowers in-migration of college-educated homeowners by about 6 percent of its pre-period mean and raises newly requested H-1B workers by about 20 percent.

A CZ-occupation-year panel shows that the H-1B response is larger where a high wedge meets strong occupation-specific labor demand. $MPW^{in}_d$ is the difference between the destination's current mortgage payment and the average old payment still held by homeowners in feeder origins. A variance decomposition shows that most of the cross-CZ spread comes from the feeder-origin component. When the two components enter the regressions separately, the migration response loads on the feeder-origin payment, while the H-1B response loads on both.

To address the concern that the feeder-origin payments are correlated with unobserved local shocks, we instrument the feeder-origin component with a predicted analog from gravity-based migration weights and lender-predicted origin payments. The IV estimates are similar to the baseline estimates. Event-study estimates place the main post-shock divergence in 2023 and 2024, and the results are robust to excluding the pandemic years, dropping the largest H-1B commuting zones, adding house-price controls, and estimating with alternative functional forms. A network-permutation placebo that randomly reassigns feeder-origin exposure across destinations leaves the baseline estimates far in the tails of the placebo distribution, and an IPEDS placebo using U.S. nonresident student enrollment shows no comparable post-2022 response, so the wedge does not capture a broader post-pandemic recovery in international inflows or a college-town concentration effect.

We also report a back-of-the-envelope calculation to put the magnitudes of the migration and H-1B estimates on a common scale. Our estimates imply 14 additional H-1B sponsorship requests and 3 approved H-1B workers per 100 deterred college-educated in-migrants. These calculations suggest that H-1B sponsorship is one of several institutional adjustment margins that U.S. firms use when domestic high-skilled in-migration becomes less responsive.

\section{Related Literature} \label{sec:related_lit}

Two literatures meet in this paper. One examines how housing-market frictions, and mortgage lock-in in particular, shape household mobility \citep{brown_locked_2020, ferreira_housing_2010, fonseca_mortgage_2024, fonseca_unlocking_2024, gerardi_mortgage_2024, howard_jue_2026, howard_effects_2025, ioannides_housing_2025, liebersohn_household_2025, notowidigdo_incidence_2020}. \citet{ferreira_housing_2010} estimate that a one-dollar increase in a homeowner's monthly mortgage payment lowers two-year mobility by about 12 percent of baseline. \citet{notowidigdo_incidence_2020} estimates a concave local housing-supply curve and shows that population adjusts asymmetrically to demand shocks, with the contraction margin absorbing more of the response than wages or housing prices. \citet{olney_determinants_2024} use bilateral commuting-zone flows from 1991 to 2013 to decompose the secular decline in U.S. internal migration, and find that falling responsiveness to origin home prices, concentrated among older homeowners in expensive states, accounts for the entire trend. This literature treats lock-in as a friction on a homeowner's decision to leave a particular origin. We measure it as a friction on a destination's ability to attract movers. The in-migration mortgage-payment wedge $MPW^{in}$ aggregates origin-side lock-in across the feeder labor markets a destination historically draws from, which lets the same national rate shock generate destination-specific variation in the cost of attracting domestic high-skilled workers.

The second literature studies high-skilled immigration and the firm-side margins through which employers adjust to labor-supply constraints \citep{hanson_high-skilled_2017, fasani_economics_2020, kerr_supply_2010, kerr_skilled_2015, doran_effects_2022}.\footnote{Related work shows that high-skilled migrants are positively selected and that networks shape destination choices \citep{parey_selection_2017, becker_persecution_2024}. Firm-level evidence links skilled immigration to employment expansion, occupational reallocation, innovation, and productivity \citep{bound_recruitment_2015, kerr_immigration_2013, nathan_wider_2014, bernstein_contribution_2022}. Work on visa constraints documents offshoring, global sourcing, and other adjustment margins \citep{mahajan_immigration_2024, glennon_how_2024, clemens_effect_2022, amuedo-dorantes_settling_2019}.} \citet{mayda_effect_2018} document that the 2004 reduction in the for-profit H-1B cap cut new H-1B hiring by 16 to 26 percent and reshaped the composition of remaining sponsorship toward computer occupations and Indian-born workers, evidence that the cap binds above the demand schedule that firms face. \citet{peri_stem_2015} estimate that foreign STEM inflows account for 30 to 50 percent of total factor productivity growth in the United States between 1990 and 2010, so the marginal H-1B worker carries productivity content. The post-2021 mortgage-rate increase that drives the variation we exploit is itself a national monetary-transmission event \citep{romer_new_2004, nakamura_identification_2018, carlino_differential_1998, beraja_regional_2019, aastveit_asymmetric_2022}, and the internal-migration channel that ordinarily equilibrates such shocks across regions \citep{blanchard_regional_1992} has weakened substantially \citep{olney_determinants_2024}. We connect the two literatures by showing that H-1B sponsorship rises in destinations where a housing-induced friction makes domestic high-skilled in-migration less responsive. The firm-side adjustment margin we identify complements the household-side mobility friction the housing-lock-in literature studies.

\section{Data and Institutional Background} \label{sec_data}

Our empirical analysis uses an annual commuting-zone panel covering 2017 to 2024. We also extend this panel to commuting-zone-by-SOC-by-year cells to study how occupation-specific labor demand shapes the H-1B response.

\subsection{H-1B Labor Condition Applications}

We use public disclosure files from the Labor Condition Application (LCA) program administered by the U.S.\ Department of Labor's Office of Foreign Labor Certification.\footnote{The OFLC LCA public disclosure files are available from the U.S.\ Department of Labor Performance Data portal: \url{https://www.dol.gov/agencies/eta/foreign-labor/performance}. Institutional background on the H-1B attestation-based programs and the LCA requirement is available at \url{https://www.dol.gov/agencies/eta/foreign-labor/programs/h-1b}. The LCA filing requirements are codified in 20 CFR 655.730: \url{https://www.law.cornell.edu/cfr/text/20/655.730}.} An LCA is an employer attestation that the Department of Labor must certify before a firm can petition USCIS for an H-1B worker. The files record application status, requested employment start dates, worksite locations, occupations, employers, and the number of requested worker positions.

Because certification precedes any USCIS petition, LCAs offer an early and geographically detailed view of intended H-1B sponsorship.\footnote{LCAs cannot be submitted more than six months before the requested beginning date.} They do not guarantee realized employment. A certified LCA may never lead to a filed petition, the petition may be denied, and an approved petition need not produce an observed job start within the same calendar year. We therefore use LCAs as measures of employer sponsorship demand rather than realized hires.

The sample spans 2017 to 2024 and consists of LCAs with certified status in the H-1B visa class.\footnote{We retain records with certified status and the H-1B visa class only, dropping denied, withdrawn, and certified-withdrawn statuses and records assigned to the related E-3 and H-1B1 visa classes.} Requested worker positions are split across six employment-action categories, and our analysis uses the new-employment count and the cross-category total.\footnote{Other categories are continued employment with the same employer, change in previously approved employment, new concurrent employment, change of employer, and amended petitions.} In total, the certified files contain over 8 million requested workers across about 4.5 million LCAs. Each LCA is dated by the calendar year of its requested employment start. Figure~\ref{fig:h1b_time} plots annual totals for requested and newly requested H-1B workers. The full counts exceed the statutory cap on new H-1B admissions because certified LCAs include continuations, amendments, and changes of employer, and because not all petitions are cap-subject.\footnote{The H-1B statutory cap is 65,000 new visas per fiscal year plus 20,000 for U.S.\ master's-degree beneficiaries (8 U.S.C.\ \S\ 1184(g)). Continuations and petitions from cap-exempt employers (mainly higher-education and nonprofit research) do not count.} 

We assign each LCA to a commuting zone by the employer-reported worksite, using the primary worksite when several are listed. The worksite ZIP is mapped to a ZIP Code Tabulation Area, the ZCTA to the county holding the largest share of its population, and the county to a 1990 commuting zone via the Dorn county-to-CZ crosswalk.\footnote{The ZCTA-to-county mapping uses the 2010 Census ZCTA-to-county relationship file (\url{https://www2.census.gov/geo/docs/maps-data/data/rel/}), and records whose ZIPs do not map to a 2010 ZCTA are dropped. The county-to-CZ crosswalk is from \citet{tolbert_us_1996} (\url{https://www.ddorn.net/data.htm}). In the full certified disclosure data, 94.9 percent of application rows map from ZIP to county, and 99.4 percent of mapped counties map to a commuting zone.} Figure~\ref{fig:h1b_map} displays the resulting CZ-level distribution of requested H-1B workers per 1,000 employed workers.

We use four CZ-year outcomes in our regressions. Newly requested workers count positions in the new-employment category. Total requested workers sum positions across all six employment-action categories. New LCAs are applications with at least one new-employment position. New filing firms count distinct firms with at least one new-employment LCA in the CZ-year.

To examine whether the H-1B response varies with occupation-specific labor demand within a commuting zone, we also build a balanced CZ-SOC-year panel. LCA occupations are aggregated to broad two-digit SOC groups, with seven H-1B-relevant groups (management, business and financial operations, computer and mathematical, architecture and engineering, life/physical/social sciences, education, and healthcare practitioners and technical occupations) retained separately and the remainder pooled into an other category. 

\subsection{Downstream H-1B Petition Outcomes From USCIS}

We augment the LCA data with USCIS administrative records on H-1B petitions and approvals to track employer sponsorship activity at a later stage of the H-1B process. Petition adjudication occurs after Department of Labor certification, so the USCIS data isolate the subset of certified LCAs that employers actually pursued through the federal petition system. A new-employment approval is the USCIS decision to grant the petition and allow the foreign individual to work. It does not record the worker's actual start of employment.

We again use four CZ-year outcomes that mirror the four LCA outcomes. These are newly approved workers, total approved workers, new petitions, and filing firms. Newly approved workers count approvals in the new-employment category, while total approved workers aggregate approvals across all H-1B employment-action categories. New petitions equal new-employment approvals plus new-employment denials. Filing firms count distinct employers with at least one adjudicated H-1B petition in the CZ-year.\footnote{The USCIS firm count is across all H-1B petition types, whereas the LCA-side new filing firms variable is restricted to new-employment LCAs.} All four enter the analysis scaled per 1,000 employed workers.

The USCIS data differ from the LCA data in two ways. First, the reporting year is the federal fiscal year of adjudication, which runs from October of the previous calendar year through September of the labeled year, rather than the calendar year of the requested employment start. Second, the geographic identifier is the petitioning employer's address rather than the intended worksite. Both differences inject measurement error into the USCIS outcomes relative to their LCA counterparts.

\subsection{The In-Migration Mortgage-Payment Wedge}
\label{sec:mpw}

Because most conventional U.S.\ mortgages are not assumable or portable, a move forces a homeowner to pay off the existing loan and finance the next purchase at the prevailing rate. When rates have risen, this swap is costly, and the cost falls on the act of moving itself, generating a wedge between old and new mortgage payments. Figure~\ref{fig:mortgage_rates} traces the national mortgage-rate cycle between 2017 and 2024 using the Freddie Mac series from FRED.\footnote{\href{https://fred.stlouisfed.org/series/MORTGAGE30US}{30-Year Fixed Rate Mortgage Average in the United States [MORTGAGE30US]}, retrieved from FRED, Federal Reserve Bank of St.\ Louis. Accessed January 25, 2026.} The post-2022 surge in rates created this wedge for homeowners with low-rate existing mortgages. 

Although the rate increase was national, the wedge it implies is geographically uneven. A move into destination $d$ depends on two payment objects, the payment a mover would face on a new mortgage in $d$ and the payment associated with the low-rate mortgage the mover would give up in the origin location $o$. Because destinations draw movers from different historical origin markets, the same national shock translates into different destination-level exposure.

We capture this exposure with the \emph{in-migration mortgage-payment wedge}, denoted $MPW^{in}_d$. The measure is the migration-weighted monthly mortgage-payment increase faced by domestic movers entering destination commuting zone $d$, where the weights are based on historical origin shares. Let $\omega_{od}$ denote the share of $d$'s 2015 to 2019 in-migrants from origin commuting zone $o$, computed from pre-shock IRS migrant counts. Let $P^{new}_d$ denote the counterfactual payment that 2020 and 2021 borrowers in destination $d$ would face at current rates, and let $P^{old}_o$ denote the payment 2020 and 2021 borrowers in origin $o$ actually pay. For a move from $o$ to $d$, the normalized monthly payment increase is
\[
\Delta P_{o \to d}
=
P^{new}_d - P^{old}_o .
\]
The in-migration mortgage-payment wedge is
\begin{equation}
\label{eq:mpw_in}
MPW^{in}_d
=
\sum_{o \neq d} \omega_{od} \Delta P_{o \to d}
=
\sum_{o \neq d} \omega_{od}
\left(P^{new}_d - P^{old}_o\right).
\end{equation}
All payments are expressed in dollars per month per \$100{,}000 of mortgage principal. This normalization makes the measure comparable across commuting zones and prevents it from mechanically reflecting differences in loan balances or home values.

The payment components of equation \ref{eq:mpw_in} come from Home Mortgage Disclosure Act mortgage originations.\footnote{HMDA is administered through the FFIEC and CFPB HMDA platforms. See \url{https://www.ffiec.gov/data/hmda} and \url{https://www.consumerfinance.gov/data-research/hmda/}.} The HMDA sample includes originated, owner-occupied, first-lien, fixed-rate, fully amortizing mortgages with approximately 30-year terms, defined as 330 to 390 months.\footnote{The sample includes both purchase and refinance originations. We exclude mortgages with balloon payments, interest-only periods, negative amortization, or other non-amortizing features. We drop observations with missing interest rates and use the HMDA-reported contract rate.} Using 2024 originations, we estimate the pricing equation
$r_{i,2024} = \alpha + X_i'\beta + \gamma_{c(i)} + \varepsilon_i$, where $r_{i,2024}$ is the annual contract interest rate, $X_i$ includes observable borrower and loan characteristics, and $\gamma_{c(i)}$ are county fixed effects. The covariates include loan size, property value, borrower income, debt-to-income ratio, discount points, lender credits, origination charges, total loan costs, loan purpose, and borrower demographic indicators.\footnote{We estimate the pricing model using ridge regression for numerical stability with a large set of county fixed effects. Results are unchanged under ordinary least squares.} Applying this 2024 pricing rule to each commuting zone's 2020 and 2021 borrowers yields $P^{new}_d$, the counterfactual payment those borrowers would face today, holding their loan and borrower characteristics fixed. The actual normalized monthly payment on the same 2020 and 2021 loans gives $P^{old}_o$.\footnote{Payments are principal-and-interest payments computed using the standard fixed-rate amortization formula, normalized to a \$100{,}000 principal amount, with $n=360$. Taxes and insurance are excluded.}

The migration-network weights $\omega_{od}$ are constructed from IRS county-to-county migration files for 2015 to 2019, aggregated to commuting zones.\footnote{IRS SOI migration data are based on year-to-year address changes reported on individual income tax returns. See \url{https://www.irs.gov/statistics/soi-tax-stats-migration-data}. The IRS aggregates or suppresses some small bilateral flows. Commuting zones with zero observed inflows after these restrictions have undefined in-migration wedges and are coded as missing.} Flows are measured by the count of taxpayers and dependents on returns that changed county of address. The IRS migration data enter the analysis only through these weights. Because the weights predate the rate shock, variation in $MPW^{in}_d$ reflects pre-existing migration networks rather than post-shock changes in flows.

\subsection{Interpreting the In-Migration Mortgage-Payment Wedge}
\label{sec:mpw_decomposition}

Because $P^{new}_d$, the destination current mortgage payment in 2024, does not vary across origins for a given destination, equation~\eqref{eq:mpw_in} collapses to
\begin{equation}
\label{eq:mpw_wop}
MPW^{in}_d
=
P^{new}_d - WOP_d,
\qquad
WOP_d
=
\sum_{o \neq d} \omega_{od} P^{old}_o ,
\end{equation}
where $WOP_d$ is the Weighted-Origin Old Payment for destination $d$. The wedge is large when the current payment environment in destination $d$ sits above the old payment environment of its feeder origins.

Figure~\ref{fig:inmigration_cost} maps $MPW^{in}_d$ across the 686 commuting zones for which we can compute the measure (out of 741).\footnote{White areas in Figure~\ref{fig:inmigration_cost} indicate commuting zones with missing values due to IRS suppression or aggregation of small flows.} The geography is not a map of expensive destinations. A high-payment destination can carry a modest wedge if its origin markets are themselves high-payment. A moderately priced destination can carry a large wedge if its movers historically came from low-payment places. Figure~\ref{fig:map_costin_components} illustrates this point by mapping the two right-hand-side terms of equation~\eqref{eq:mpw_wop}, the destination current payment $P^{new}_d$ and the weighted-origin old payment $WOP_d$, separately.

The two components contribute unequally to the cross-commuting-zone variance of the mortgage wedge, $MPW^{in}$. The variance of $MPW^{in}$ is 49.3, measured in squared monthly dollars per \$100{,}000 of principal. The variance of $P^{new}$ is 9.0 and the variance of $WOP$ is 46.9. The covariance term contributes $-6.6$, and the correlation between $P^{new}$ and $WOP$ is only 0.16. Most of the cross-commuting-zone variation in $MPW^{in}$ comes from where movers came from, not from where they are moving to.

$MPW^{in}_d$ is a CZ-level exposure measure, not an individual-level moving cost. Coefficients on $MPW^{in}_d$ are reduced-form effects of this exposure, not moving-cost elasticities. The standard deviation of $MPW^{in}$ is approximately \$7.02 per month per \$100{,}000 of principal. For a \$300{,}000 mortgage, a one-standard-deviation increase corresponds to about \$21 per month, or \$252 per year. Discounted at 5 percent, this stream has a present value of approximately \$3{,}900 over 30 years and about \$2{,}000 over 10 years.\footnote{The present value of a constant monthly wedge $A$ sustained for $T$ years and discounted at annual rate $\delta$ is $A \cdot \dfrac{1-(1+\delta/12)^{-12T}}{\delta/12}$.}

\subsection{Other Variables}\label{sec:other_vars}

The ACS 1-year Public Use Microdata Samples supply several of the other variables we use in the analysis.\footnote{We assign respondents to CZs from their state and Public Use Microdata Area (PUMA) of residence using the area-factor weights of \citet{dorn_essays_2009}, with 2010 PUMA definitions for ACS 2017 to 2021 and 2020 definitions for 2022 to 2024. Migration origins are mapped analogously through the ACS migration PUMA. See \url{https://www.census.gov/programs-surveys/acs/microdata.html}.} These include demographic and socioeconomic controls (age structure, race and ethnicity shares, foreign-born share, home ownership, and college-plus and STEM shares), as well as the migration outcomes, both domestic in-migration (by education and homeowner-renter status) and inflows from abroad.

Our regressions also control for the teleworkable share, defined as the employment-weighted CZ-level share of jobs in occupations that can be performed from home, based on the \citet{dingel_how_2020} classification.\footnote{See \url{https://github.com/jdingel/DingelNeiman-workathome/blob/master/onet_to_BLS_crosswalk/output/onet_teleworkable_blscodes.csv}.} County unemployment and employment come from the BLS Local Area Unemployment Statistics and Quarterly Census of Employment and Wages, aggregated to commuting zones.\footnote{See \url{https://www.bls.gov/lau/} and \url{https://www.bls.gov/cew/}.} Baseline technology intensity, used in the heterogeneity analysis, is the 2017 to 2019 CZ employment share in seven 4-digit NAICS technology industries.\footnote{Software publishers (5112), other telecommunications (5179), data processing and hosting (5182), other information services (5191), computer systems design (5415), scientific research and development (5417), and other professional, scientific, and technical services (5419).}

For the occupation-demand analysis, we construct a SOC-specific Bartik shock from ACS employment. The shock is the product of each CZ's 2017 to 2019 baseline SOC employment share and national log employment growth in that group relative to the baseline mean.

Using the Federal Housing Finance Agency's House Price Index, we construct log housing prices and annual percent changes for the robustness checks.\footnote{FHFA HPI data are at \url{https://www.fhfa.gov/data/hpi/datasets}.} For a placebo analysis, IPEDS Fall Enrollment supplies counts of foreign students at Title IV institutions, which we aggregate to commuting zones, along with undergraduate and graduate breakouts, total enrollment, and the foreign share.\footnote{See \url{https://nces.ed.gov/ipeds/survey-components/8}. Foreign students are reported under the \emph{U.S.\ Nonresident} category. For the terminology change, see \url{https://nces.ed.gov/ipeds/survey-components/release-memo?type=fall&year=2023}.}

\subsection{Summary Statistics}\label{sec:sumstats}

Table~\ref{tab:sumstats} reports summary statistics for the estimation sample. Panel A summarizes the migration-weighted lock-in cost measures (in \$/month per \$100{,}000 of mortgage principal). Panel B reports the distribution of requested H-1B worker positions from certified LCAs, both in total and by employment-action category. Panels C and D summarize the demographic and industry controls used in the empirical analysis.

\section{Empirical Analysis} \label{sec_empirical}

\subsection{Identification}
\label{sec:identification}

The baseline design is a continuous-treatment difference-in-differences exposure design. For commuting zone $c$ in year $t$, we estimate
\begin{equation}
\label{eq:baseline}
Y_{ct}
=
\beta
\left(
MPW^{in}_c \times Post_t
\right)
+
X_{ct}'\gamma
+
\alpha_c
+
\delta_t
+
\varepsilon_{ct},
\end{equation}
where $Y_{ct}$ is a migration or H-1B outcome, $MPW^{in}_c$ is the wedge defined in equation~\eqref{eq:mpw_wop}, $Post_t$ equals one for 2022 to 2024, $X_{ct}$ is a vector of time-varying commuting-zone controls, and $\alpha_c$ and $\delta_t$ are commuting-zone and year fixed effects. Standard errors are clustered at the commuting-zone level. 

The identifying variation in $MPW^{in}_c$ comes from differential exposure of commuting zones to the common national rate increase. Destination commuting zones differ in the origin markets from which they historically draw movers, and those origins differ in the old mortgage-payment environments that potential movers would have to leave behind. The IRS migration weights are measured before the rate shock, using 2015 to 2019 flows, and the outcomes do not enter the construction of $MPW^{in}_c$.

The causal effects can be recovered under the assumption that, conditional on commuting-zone and year fixed effects and time-varying controls, high- and low-$MPW^{in}$ commuting zones would have followed similar post-2022 trends in domestic high-skilled in-migration and H-1B sponsorship, absent the mortgage-lock-in channel. Equivalently, the mortgage payments embedded in a destination's pre-shock feeder-origin network must not be correlated with post-2022 shocks to migration or employer sponsorship through channels other than the moving-cost wedge created by mortgage lock-in. Because $MPW^{in}_c = P^{new}_c - WOP_c$, and most of the cross-CZ variation in the wedge comes from $WOP_c$, this assumption is primarily about the weighted-origin old-payment component rather than about destination payment levels alone.

Several threats may violate this assumption. For example, origin markets may differ in ways correlated with post-2022 labor-demand shocks, through persistent industry, occupation, or recruiting links, or through the 2020 to 2021 credit conditions that shaped their old mortgage payments. Destination current payments may instead proxy for expensive or fast-growing places on different trends. Post-pandemic international inflows could also recover in high-wedge destinations for reasons unrelated to mortgage lock-in. We address these issues with a number of supplementary analyses and robustness checks. 

\subsection{Domestic and International In-migration Response}
\label{sec_migration_response}

Table~\ref{tab:migration_costin} presents estimates of equation~\eqref{eq:baseline} for migration per 1,000 residents. Panel A covers domestic in-migration. Higher-wedge destinations receive fewer domestic in-migrants after 2022 across most groups, and the response is largest among college-educated homeowners. The estimates for total college in-migration (column 2) and college-owner in-migration (column 4) are identical to three decimal places, while the college-renter estimate (column 5) is more than an order of magnitude smaller in absolute value and statistically indistinguishable from zero. Homeowners are directly exposed to the wedge because relocation requires retiring an existing low-rate mortgage and originating a new loan at prevailing rates, while renters are not. The owner-renter asymmetry is consistent with mortgage lock-in operating through the channel implied by the wedge. \citet{olney_determinants_2024} document the same homeowner-renter contrast in 1991 to 2013 bilateral migration data, where the long-run decline in mobility is concentrated among older homeowners and absent for renters.

These estimates imply that a one-standard-deviation increase in $MPW^{in}$ (\$7.02 per month per \$100{,}000 of principal) lowers college-owner in-migration by 0.41 movers per 1,000 residents, or about 6 percent of the pre-period mean of 7.06. The effect on total college in-migration is also 0.41 per 1,000, or 2.9 percent of its pre-period mean. 

Figure~\ref{fig:event_college_migration} reports the event-study coefficients for college in-migration, relative to 2019. The 2017 and 2018 coefficients are near zero, the 2020 coefficient is negative (reflecting pandemic-era mobility disruptions), and the post-shock coefficients become more negative in 2023 and 2024, with confidence intervals that exclude zero.

Panel B of Table~\ref{tab:migration_costin} turns to inflows from abroad. The ACS migration data identify people who lived abroad in the previous year and now reside in the destination commuting zone, but they do not record specific visa status (for example, an F-1 student visa, an H-1B, or permanent residency). The estimates should therefore be interpreted as realized high-skilled inflows. Column 1 shows that college-educated inflows rise in high-wedge destinations after 2022. The coefficient on $MPW^{in}_c \times Post_t$ is $0.015$ and is significant at the 1 percent level. In contrast, the corresponding non-college coefficient in column 2 is small and statistically insignificant. The same destinations that receive fewer domestic college-educated movers thus take in more college-educated inflows from abroad.

Columns 3 to 6 split the foreign inflows by citizenship. The college-educated increase comes almost entirely from non-citizens. The coefficient on $MPW^{in}_c \times Post_t$ is a significant $0.013$ for college-educated non-citizen inflows (column 4), while the estimate for college-educated citizen inflows is virtually zero (column 3). Non-college non-citizen inflows decline modestly (column 6). This non-citizen pattern is what one would expect if the foreign inflow into high-wedge destinations operates through visa-based channels, because citizens arriving from abroad do not require employer sponsorship.

Table~\ref{tab:migration_decomposition} examines the level decomposition from Section~\ref{sec:mpw_decomposition}, which splits the wedge into the 2024 Mortgage Payment $P^{new}_c$ (the destination's counterfactual payment at 2024 rates) and the Locked-In Origin Payment $WOP_c$ (the migration-weighted actual 2020 to 2021 payment in origin markets). For the college-owner outcome in column 4, the 2024 Mortgage Payment enters negatively, because a higher destination payment raises the cost of moving in, and the Locked-In Origin Payment enters positively, because a higher old payment in feeder origins leaves potential movers with less low-rate mortgage value to surrender. The two coefficients, $-0.060$ and $0.058$, are nearly mirror images, and a test that they are exactly equal and opposite, $\beta_P + \beta_W = 0$, does not reject (p = 0.96), so the migration response is well summarized by the wedge itself. The cross-CZ variation the design exploits, however, comes overwhelmingly from the feeder-origin component. Because $WOP_c$ has far more cross-CZ dispersion than $P^{new}_c$, a one-standard-deviation change in $WOP_c$ shifts college-owner in-migration by about 0.40 per 1,000 residents, against about 0.18 for a one-standard-deviation change in $P^{new}_c$. High-wedge destinations therefore lose domestic college-owner in-migrants mainly when their historical feeder origins contain homeowners holding especially valuable low-rate mortgages.

Appendix Figure~\ref{fig:event_mig_components} presents event studies using the college in-migration rate as the outcome on the two components and confirms the post-2022 divergence in the Locked-In Origin Payment, while the 2024 Mortgage Payment coefficients remain near zero.

\subsection{H-1B Activity at the Commuting-Zone Level}
\label{sec_h1b_cz}

The migration results show that high-wedge destinations receive fewer domestic college-owner in-migrants after 2022, even as the ACS records more college-educated arrivals from abroad. The ACS foreign-inflow results, however, do not identify visa status. As a result, we turn to administrative H-1B records to test whether employers increased formal sponsorship activity after the mortgage-rate shock. To investigate, we estimate equation~\eqref{eq:baseline} with H-1B outcomes per 1,000 employed workers as the dependent variable. These outcomes come from certified Labor Condition Applications (LCA) and consist of newly requested workers, total requested workers, new LCAs, and firms filing new-employment LCAs, all of which measure employer sponsorship demand rather than realized hires.

Table~\ref{tab:h1b_cz} reports the estimates. Employer demand for H-1B sponsorship rises in high-wedge destinations after 2022 on all four LCA margins, with each estimate significant at the 1 percent level. The firm-count response (column 4) indicates that the increase reflects more distinct firms filing new-employment LCAs in high-wedge destinations rather than larger petitions from incumbent sponsors.\footnote{\citet{mayda_effect_2018} find that the binding cap concentrates realized H-1B issuances among firms that already use the program intensively. We measure firm counts at the LCA-filing stage, which sits upstream of the cap, so our response captures the demand-side margin rather than the cap-induced rationing outcome.}

Scaling these coefficients by the standard deviation of $MPW^{in}$ yields implied responses of about 0.13 newly requested H-1B workers and 0.015 new filing firms per 1,000 employed workers (roughly 20.0 and 9.8 percent of their pre-period means). Figure~\ref{fig:event_h1b_workers_new_pe} plots the year-by-year coefficients on newly requested H-1B workers with 2019 omitted. The pre-2022 coefficients are positive but imprecisely estimated, while the post-shock estimates rise after 2022 and remain elevated through 2024, consistent with a post-2022 increase in H-1B sponsorship demand in the same high-wedge destinations where domestic college-owner in-migration weakens.

Table~\ref{tab:h1b_decomposition} applies the same decomposition to H-1B sponsorship (Table~\ref{tab:migration_decomposition}). If sponsorship responds to the wedge, the 2024 Mortgage Payment should enter positively and the Locked-In Origin Payment negatively. For newly requested workers, our main H-1B outcome, both components carry the predicted sign and are statistically significant. A higher destination payment makes recruiting domestic movers more costly and raises employer demand for foreign sponsorship, while a higher old payment in feeder origins leaves origin markets with less old-rate advantage to lose. The same sign pattern holds across the other LCA margins.

Unlike the migration response, which is primarily driven by the Locked-In Origin Payment component, the H-1B response loads on both. Scaled by their cross-CZ standard deviations, $P^{new}_c$ and $WOP_c$ contribute about equally to newly requested workers, roughly 0.11 per 1,000 employed for each. That is, H-1B sponsorship requests rise both where entering the destination is costly under current pricing and where mortgage lock-in is severe in the origins that feed it. Appendix Figure~\ref{fig:event_h1b_components} plots year-by-year coefficients for the two components on newly requested workers and shows the 2024 Mortgage Payment turning positive and the Locked-In Origin Payment negative after the 2022 shock.

\subsection{Occupation-Specific Demand and H-1B Activity}
\label{sec_h1b_cz_soc}

The CZ-year results establish that H-1B sponsorship rises in high-wedge destinations after 2022. If a mobility friction lies behind that rise, the effect should be sharpest in occupations where hiring demand is most acute. As an initial and coarse test, we divide commuting zones by baseline technology-employment intensity, and Appendix Table~\ref{tab:h1b_heterogeneity_tech} shows a weaker worker-request response in the most technology-intensive of them, in keeping with the slowdown in technology hiring over the same period. Because that division blends many occupations together with confounding local conditions, we instead adopt a CZ-SOC-year design that isolates the occupation-demand margin within a commuting zone and year.

We estimate
\begin{equation}
\label{eq_cz_soc_year}
\begin{split}
H_{cst}
&=
\beta
\left(
MPW^{in}_c
\times
Post_t
\times
\widetilde{B}_{cst}
\right) \\
&\quad
+
\pi_1 \widetilde{B}_{cst}
+
\pi_2
\left(
Post_t
\times
\widetilde{B}_{cst}
\right)
+
\pi_3
\left(
MPW^{in}_c
\times
\widetilde{B}_{cst}
\right) \\
&\quad
+
\alpha_{ct}
+
\alpha_{cs}
+
\alpha_{st}
+
\varepsilon_{cst},
\end{split}
\end{equation}
where $H_{cst}$ is H-1B activity in commuting zone $c$, SOC group $s$, and year $t$ (per 1,000 employed workers in the CZ-SOC cell).\footnote{The eight SOC groups are (1) management, (2) business and financial operations, (3) computer and mathematical, (4) architecture and engineering, (5) life, physical, and social sciences, (6) education, (7) healthcare practitioners and technical, and (8) an ``Other'' residual category combining the remaining occupations. See Section~\ref{sec_data} for the aggregation.} $\widetilde{B}_{cst}$ is the demeaned SOC-specific Bartik demand shock in percentage points, built by interacting national SOC employment growth with each CZ's pre-period SOC employment mix. The three sets of two-way fixed effects (CZ-by-year, CZ-by-SOC, and SOC-by-year) absorb the corresponding aggregate, occupational, and national shocks, along with the lower-order terms in $MPW^{in}_c$, $Post_t$, and $MPW^{in}_c \times Post_t$. Identification of $\beta$ comes from within-CZ-year, within-SOC-year variation in H-1B activity across occupations with different predicted demand growth in commuting zones with different wedge exposure. Standard errors are clustered at the commuting-zone level.

The triple interaction in Table~\ref{tab:h1b_cz_soc_bartik} is significant at the 1 percent level for newly requested workers, total requested workers, and new LCAs (columns 1 through 3), while the new-filing-firm estimate in column 4 is positive but not statistically distinguishable from zero. Multiplying these coefficients by the standard deviation of $MPW^{in}$ ($\$7.02$ per month per \$100{,}000 of principal) yields incremental responses, per one-percentage-point increase in the SOC Bartik shock, of 0.337 newly requested workers per 1,000 employed workers. This effect corresponds to 11.2 percent of the pre-period mean. Within a high-wedge destination, then, the increase in H-1B sponsorship is largest in occupations with the strongest demand growth.

Appendix Table~\ref{tab:h1b_cz_soc_bartik_decomposition} replaces the wedge in the triple interaction with the 2024 Mortgage Payment and the Locked-In Origin Payment. The 2024 Mortgage Payment is positive and significant on the two worker margins, and the Locked-In Origin Payment is negative and significant there and on new LCAs, as the wedge mechanism implies. Because the Locked-In Origin Payment varies far more across commuting zones, it accounts for the larger part of the amplified response on the worker and LCA margins, while the 2024 Mortgage Payment matters more only for firm counts. The occupation-demand amplification therefore runs mainly through feeder-origin lock-in and is not simply an expensive-destination effect. The pattern is strongest in high-demand occupations where mortgage lock-in is most likely to constrain the domestic in-migration margin.

A binned version of the test, in Appendix Table~\ref{tab:positive_bartik_bins}, shows the response confined to cells with strongly positive predicted demand, with nothing comparable in low-positive cells. Figure~\ref{fig:leaveoneout_soc} re-estimates the specification, dropping one SOC group at a time, and the triple-interaction coefficient stays positive in every case, including with computer and mathematical occupations removed.

\subsection{Downstream H-1B Petition and Approval Outcomes}
\label{sec_uscis_downstream}

The LCA results measure employer requests for H-1B sponsorship and do not capture petition adjudication or employment starts. The same post-2022 pattern appears in USCIS administrative records on H-1B petitions and approvals, which sit one step closer to completed sponsorship. The two data sources also differ in reporting conventions. USCIS publishes its records only by fiscal year of petition adjudication and at the petitioning employer's address (the only fields the agency releases), whereas LCAs are reported by requested employment start year and at the intended worksite. These differences attenuate any tight mapping between the two, so the USCIS estimates are best read as downstream corroboration rather than a one-for-one replication of the LCA results. Because approvals still do not record actual job starts, we interpret the USCIS outcomes as evidence of completed sponsorship approvals rather than realized employment.

Table~\ref{tab:uscis} estimates equation~\eqref{eq:baseline} using the four USCIS outcomes. All four coefficients carry the same sign as the LCA estimates and are smaller in magnitude, consistent with USCIS sitting downstream of LCA filings and excluding requests that never reach approval. The new-filing-firm margin (column 4) is significant at the 1 percent level, and the worker and petition margins are significant at the 5 or 10 percent level. The positive USCIS coefficients support the interpretation that the LCA results capture genuine post-2022 movement in H-1B sponsorship rather than speculative or unused filings.

Appendix Table~\ref{tab:uscis_decomposition} applies the level decomposition to the USCIS outcomes. Most coefficients are imprecise, with the filing-firm column the most clear exception, where the 2024 Mortgage Payment is positive at the 10 percent level and the Locked-In Origin Payment is negative at the 1 percent level. In that column the larger contribution falls on the Locked-In Origin Payment, which varies far more across commuting zones than the destination payment, consistent with the lock-in channel behind the migration results. The USCIS decomposition is noisier than its LCA counterpart, where both components are estimated more precisely, but the signs line up with the same underlying mechanism.

\subsection{Instrumental-Variables Strategy}
\label{sec:iv}

Section~\ref{sec:identification} flagged that the identifying variation in $MPW^{in}_c$ is concentrated in the locked-in origin payment, so the leading concern is that a destination's feeder origins are tied to industries or networks with distinct post-2022 high-skilled labor-demand shocks. If so, the H-1B response could reflect those shocks rather than the mortgage lock-in channel. The instrumental-variables exercise is designed to reduce this concern by replacing realized feeder-origin exposure with predicted exposure based on pre-shock geography, population, and lender-pricing components. The exclusion restriction is that these predicted components affect post-2022 migration and H-1B sponsorship only through their effect on mortgage-lock-in exposure, conditional on the fixed effects and controls.

\subsubsection{Constructing the Instrument}

To address this concern, we instrument the locked-in origin payment, $WOP_d=\sum_{o \neq d} \omega_{od} P^{old}_o$, the inner product of pre-shock migration-network weights $\omega_{od}$ and origin payments $P^{old}_o$. The instrument replaces both the realized weights and the realized payments with predicted analogs.

The first substitution replaces realized origin payments with lender-predicted payments. For each county, we combine the county's pre-shock lender shares with each lender's pricing position estimated from its 2020--2021 originations outside the county's state.\footnote{The lender-predicted origin payment is constructed in five steps. (1) From 2018--2019 HMDA originations meeting our baseline sample restrictions, we compute lender shares $s_{c,L}$ at the county level. (2) On 2020--2021 originations meeting the same restrictions, we estimate the pooled regression $p_i = X_i'\beta + \mu_c + \kappa_{s,t} + \lambda_{L,s} + u_i$, where $p_i$ is the normalized monthly payment per \$100{,}000, $X_i$ contains the borrower and loan controls used in the baseline pricing model, and $\mu_c$, $\kappa_{s,t}$, and $\lambda_{L,s}$ are county, state-year, and lender-by-state fixed effects. (3) We extract the lender-by-state fixed effects $\lambda_{L,s}$ and normalize them within state to loan-weighted mean zero. (4) For each lender $L$ and leave-out state $k$, we compute the loan-count-weighted average of $\lambda_{L,s}$ across states $s \neq k$, $q_{L,-k} = \sum_{s \neq k} n_{L,s}\,\lambda_{L,s} \big/ \sum_{s \neq k} n_{L,s}$, subject to a minimum of 50 out-of-state loans. (5) The county-level prediction is $\widehat{P^{old}}_c = \sum_L s_{c,L}\, q_{L,-s(c)}$, retained for counties where lenders with non-missing $q$ cover at least 70 percent of pre-shock loans. County predictions are then aggregated to origin commuting zones using 2018--2019 loan-count weights.} The pricing positions therefore reflect each lender's national cost of capital, securitization channels, and underwriting standards rather than local mortgage-market conditions in the county. Because the leave-out construction excludes loans originated in the county's own state, the predicted payment cannot reflect labor-market conditions in any one destination commuting zone.

The second substitution replaces realized migration-network weights with gravity-predicted weights. A gravity model predicts bilateral flows between two places from the size of the origin, the size of the destination, and the distance between them. We estimate a log-linear gravity model on pre-shock commuting-zone-to-commuting-zone migration flows and use the predicted shares in place of the realized IRS shares.\footnote{We estimate $\log(\text{flow}_{od}) = \beta_0 + \beta_1 \log(\text{pop}_o) + \beta_2 \log(\text{pop}_d) + \beta_3 \log(\text{dist}_{od}) + u_{od}$ by OLS on the cross-section of commuting-zone pairs $(o, d)$ with $o \neq d$. The dependent variable is the 2015--2019 average annual IRS county-to-county migration flow, aggregated to commuting-zone pairs. Populations come from the ACS 2019 5-year estimates. Each commuting zone's centroid is the population-weighted average of its county centroids from the 2020 Census Gazetteer, and $\text{dist}_{od}$ is the great-circle distance between origin and destination centroids. Predicted flows are exponentiated and renormalized into in-migration shares, $\hat\omega^{IV}_{od} = \widehat{\text{flow}}_{od} \big/ \sum_{o' \neq d} \widehat{\text{flow}}_{o'd}$.} The inputs to the predicted shares are pre-shock populations and the geographic distance between commuting zones. These demographic and geographic primitives remove the realized industry- and occupation-specific migration links embedded in observed feeder shares, though the exclusion restriction still requires that the predicted network affects post-2022 outcomes only through mortgage-lock-in exposure.

The predicted origin payments are then aggregated through the gravity-based feeder network into a predicted locked-in origin payment for each destination. In the instrumental-variables specification the locked-in origin payment interacted with $Post_t$ is treated as endogenous, and the destination 2024 mortgage payment interacted with $Post_t$ enters as a control.

\subsubsection{2SLS Estimates}

Table~\ref{tab:iv_wop} reports the estimates. The instrument is a strong predictor of the endogenous regressor. The first-stage coefficient is positive, and the cluster-robust Kleibergen--Paap first-stage $F$-statistic is 392.16, far above conventional weak-instrument thresholds. The gravity-based and lender-predicted construction therefore retains a strong mechanical link to the realized locked-in origin payment while discarding the realized network and realized payment variation that motivates the exercise. In the second stage, a higher locked-in origin payment raises domestic college-educated in-migration. The estimates for college in-migration and college-owner in-migration are statistically indistinguishable from each other and significant at the 1 percent level, while the college-renter estimate is close to zero and statistically insignificant. The 95 percent confidence interval for the renter group rules out increases larger than roughly 0.030 movers per 1,000 residents, well below the owner estimate. The wedge variation arising from feeder-origin lock-in therefore depresses domestic college-educated in-migration, with the response concentrated among college-owner movers.

The H-1B outcomes move in the opposite direction. The locked-in origin payment coefficient is negative for newly requested H-1B workers, new LCAs, and new filing firms, and each estimate is significant at the 1 percent level. The signs match the baseline interpretation, because a lower old-payment environment in feeder origins widens the in-migration wedge and raises H-1B sponsorship activity. The instrumental-variables estimates therefore ease the endogeneity concern raised in Section~\ref{sec:identification}.

Appendix Table~\ref{tab:iv_mpw} reports a companion specification that instruments the full in-migration mortgage-payment wedge directly, using the same gravity-based and lender-predicted components. The first stage there remains strong, with a Kleibergen--Paap $F$-statistic of 70.06, and the direct-wedge estimates point in the same direction as the baseline for college-owner in-migration, newly requested H-1B workers, and new filing firms.

\subsection{Robustness Checks and Additional Diagnostics}

The robustness checks below are organized around the main identification threats. First, we test whether the mortgage wedge is driven by noisy migration weights or by arbitrary cross-sectional variation in origin payments rather than by destination-specific feeder networks. Second, we test whether the H-1B results are driven by large technology and finance hubs, pandemic-period mobility disruptions, the 2024 H-1B rule change, or coincident destination housing-price trends. Third, we check whether the pattern reflects a broader post-2022 recovery in foreign inflows rather than employer sponsorship demand.

\subsubsection{Validity of the Mortgage Wedge Construction}

The first set of exercises addresses the validity of $MPW^{in}_c$ as a measure of lock-in costs, rather than an unmeasured destination-level shock that moves both migration and H-1B sponsorship through other channels.

Recall that $WOP_d$ averages payments across all feeder origin markets, including many small ones whose pre-shock migration weights may contain noise. Table~\ref{tab:robust_wop_topk} truncates to the top-$K$ origin markets by pre-shock inflows for $K \in \{1,3,5,10,20\}$, renormalizes the weights, and recomputes MPW$^{in,K}_d = P^{new}_d - WOP^K_d$. The coefficients converge monotonically to the baseline as $K$ grows. The top five origin markets (median 92 percent of inflows) already deliver the baseline, and even $K=1$ (median 41 percent of inflows) yields estimates significant at the 1 percent level for college in-migration and newly requested H-1B workers, at roughly two-thirds and five-sixths of the respective baseline magnitudes. The largest few origin CZs carry the identifying signal, not the long tail of small origins.

Figure~\ref{fig:robust_wop_netperm} reports results of a permutation test. For each of 1{,}000 placebo iterations, we shuffle $WOP_d$ across destinations while holding each $P^{new}_d$ fixed, preserving the marginal distributions of both objects but breaking the destination-origin match. If random origin networks reproduced the baseline, the identifying variation would come from the cross-section of $WOP$ values rather than from the destination-specific origin-market composition. The actual estimates lie far in the tails (two-sided centered $p<0.001$ for college in-migration and $p=0.011$ for newly requested H-1B workers), indicating that the destination-origin match itself is doing the identifying work.

\subsubsection{Sample Restrictions and Additional Controls}

The major technology and finance hubs account for a disproportionate share of LCA filings and might drive the headline pattern through their own idiosyncratic dynamics rather than through the wedge. Dropping the 25, 50, or 100 commuting zones with the highest H-1B activity, in Table~\ref{tab:drop_top_robustness}, barely moves the migration and H-1B estimates.

Pandemic-era mobility disruptions could create spurious correlation between $MPW^{in}_c$ and post-2022 migration or H-1B activity. Excluding the pandemic years 2020 and 2021, in Table~\ref{tab:drop_covid}, preserves the signs and significance of the baseline estimates, with magnitudes slightly larger than in the full sample.

Table~\ref{tab:drop_2024} excludes 2024 to address the changes imposed by DHS in that year.\footnote{Effective March 2024, DHS moved H-1B cap selection from registration-centric to beneficiary-centric. This rule could have altered employer filing incentives and the composition of cap-subject activity in the final year of our sample.} Excluding 2024 attenuates the H-1B coefficients by roughly one-third while leaving the migration estimates and signs intact. Newly requested workers, new LCAs, new filing firms, college-owner in-migration, and college foreign inflows all remain significant and carry the baseline signs. 

If the mortgage-payment wedge is partly capturing coincident house-price movements that the baseline controls do not absorb, the coefficients should attenuate once these controls are added. Table~\ref{tab:hpi_controls} adds two FHFA House Price Index controls, the log price level and its annual percent change. The migration estimates are essentially unchanged, while the H-1B coefficients attenuate but retain their baseline signs, with three of four columns remaining significant at conventional levels.

\subsubsection{Functional Form}

The OLS-on-count-per-1{,}000 specification could be sensitive to the functional form chosen for the count outcome. Table~\ref{tab:alt_functional_forms} re-estimates the H-1B specification under $\log(0.1+y)$, $\log(1+y)$, the inverse hyperbolic sine, conditional fixed-effects negative binomial, and a 2019-to-2024 cross-sectional long difference. All twelve coefficients across the three OLS transformations are positive and significant at the 1 percent level, the negative binomial preserves the two worker margins, and the long-difference cross-section confirms the result outside the panel structure.

\subsubsection{A Placebo Test with Foreign Students}

As a placebo, we ask whether $MPW^{in}_c$ predicts a flow that mortgage lock-in should not affect. H-1B workers and foreign college students often locate in the same commuting zones, the college towns and technology hubs, so the wedge could partly overlap with the post-2022 recovery in foreign-student enrollment. Table~\ref{tab:ipeds_placebo} replaces the migration and H-1B outcomes with IPEDS college enrollment (foreign total, foreign undergraduate, foreign graduate, total enrollment, and the foreign share). All five coefficients on $MPW^{in}_c \times$ Post are statistically zero, so the wedge does not pick up a broader post-2022 shift in foreign-population flows, and the H-1B response is not an artifact of that overlap.

\subsection{Scaling Domestic In-Migration and H-1B Sponsorship Responses}
\label{sec:offset}

To place the migration and H-1B estimates on a common scale, we report a back-of-the-envelope ratio of the H-1B sponsorship response to the decline in domestic college-educated in-migration that the same wedge increase produces. Let \(\hat\beta_M\) denote the effect of a unit increase in \(\mathrm{MPW}^{in}_c\) on college-educated in-migration per 1,000 residents, and \(\hat\theta_H\) denote the effect of the same increase on H-1B positions, petitions, or approvals per 1,000 employed workers. If \(\bar e\) is the ratio of employed workers to residents, then the relative-response ratio can be written as,

\begin{equation}
\label{eq:offset}
\mathcal{O}_H
=
\frac{\Delta H}{-\Delta M}
=
\frac{\bar e \cdot \hat\theta_H}{|\hat\beta_M|}.
\end{equation}

Our estimates indicate that \(\hat\beta_M\) is $-0.059$ (college-owner column of Table~\ref{tab:migration_costin}) and \(\hat\theta_H\) is 0.018 for newly requested H-1B workers (Table~\ref{tab:h1b_cz}). In our sample period, \(\bar e\) is about $0.45$. Thus, $\mathcal{O}^{LCA,new} = \frac{0.45 \times 0.018}{0.059} \approx 0.14$, or 14 requested H-1B positions for every 100 deterred domestic college graduates. The relative response is smaller for the USCIS-approved H-1B positions (Table~\ref{tab:uscis}), $\mathcal{O}^{USCIS,New} = \frac{0.45 \times 0.0043}{0.059} \approx 0.03$, or 3 new H-1Bs per 100 deterred domestic college graduates. \citet{peri_stem_2015} document large productivity gains from foreign STEM workers in U.S. metropolitan economies, so the welfare-equivalent substitution rate may differ from the unit-for-unit offset reported here.

\section{Summary and Conclusion} \label{sec_conclusion}

This paper asks whether mortgage lock-in in the owner-occupied housing market reshapes domestic high-skilled mobility and firms' use of the H-1B program. The 2022 rise in U.S.\ mortgage rates widened the gap between borrowers' contract rates and prevailing market rates, raising the monthly payment a household would owe if it moved and financed a new purchase. Because destinations draw workers from different feeder markets, the national rate shock generated cross-sectional variation in destination-specific moving costs. We build a destination in-migration mortgage-payment wedge from HMDA loans and pre-shock IRS migration links, and we trace its effects on domestic migration and employer demand for high-skilled foreign labor across a 2017 to 2024 panel of U.S.\ commuting zones. Event-study evidence, placebo evidence on foreign student enrollment, and an instrumental-variables design based on gravity-predicted feeder exposure support the mortgage-lock-in interpretation.

A one-standard-deviation increase in the wedge lowers college-educated homeowner in-migration by about six percent of the pre-period mean, with no detectable response among college-educated renters or among non-college workers. The owner-renter contrast indicates that the mortgage-payment wedge itself, rather than broader local shocks correlated with exposure, drives the migration response. In the same high-wedge commuting zones, employer demand for H-1B sponsorship rises by roughly twenty percent of the pre-period mean on the newly requested margin. USCIS petitions and approvals move in the same direction by smaller amounts. The sponsorship response is also larger in commuting-zone occupation cells with stronger predicted labor demand, which is where a thinner domestic in-migration margin should bind most.

Mortgage lock-in is usually framed as a friction on incumbent movers. We show that it also operates as a destination-side labor-market shock, changing which destinations can recruit the domestic workers they normally attract. When a destination's feeder markets are locked in, its domestic high-skill recruiting margin becomes less responsive to demand, and firms shift part of their adjustment toward employer-sponsored immigration. In this sense, the same mortgage contracts that insulate incumbent borrowers from rate increases transmit the rate shock to firms in other places through migration networks.

Read together, the migration and sponsorship estimates imply roughly fourteen requested H-1B positions and about three USCIS approvals per one hundred deterred domestic college-educated in-migrants. The estimated H-1B response is therefore far below a one-for-one replacement margin. H-1B sponsorship is one adjustment margin among several, and it is bounded by visa rules, employer willingness to sponsor, and the lottery for cap-subject petitions. Housing finance is therefore a channel through which monetary policy shapes employment-based immigration demand, alongside its better-known effects on residential investment and consumption.

\clearpage
\bibliography{references}

\clearpage

\section{TABLES AND FIGURES}

\input{new_figures/fig_h1b_time}

\input{new_figures/fig_mortgage_rate}

\input{new_figures/fig_inmigration_cost}

\input{new_figures/fig_map_costin_components}

\input{new_figures/fig_event_migration}

\input{new_figures/fig_event_h1b}

\clearpage
\input{new_tables/tab_sumstats}

\input{new_tables/tab_migration}

\input{new_tables/tab_mig_decomp}

\input{new_tables/tab_h1b_cz}

\input{new_tables/tab_h1b_cz_decomposition}

\input{new_tables/tab_cz_soc_h1b}

\input{new_tables/tab_uscis}

\input{new_tables/tab_iv_wop}

\clearpage
\appendix
\renewcommand{\thesection}{A\arabic{section}}
\renewcommand{\thetable}{A\arabic{table}}
\renewcommand{\thefigure}{A\arabic{figure}}
\setcounter{section}{0}
\setcounter{table}{0}
\setcounter{figure}{0}

\input{new_figures/fig_h1b_map}

\input{new_figures/fig_event_mig_components}

\input{new_figures/fig_event_h1b_cz_components}

\input{new_figures/fig_loo_soc}

\input{new_figures/fig_wop_netperm}

\clearpage

\input{new_tables/tab_h1b_tech_intensity}

\input{new_tables/tab_h1b_cz_soc_decomposition}

\input{new_tables/tab_bartik_bins}

\input{new_tables/tab_uscis_decomposition}

\input{new_tables/tab_iv_mpw}

\input{new_tables/tab_wop_topk}

\input{new_tables/tab_drop_top}

\input{new_tables/tab_drop_covid}

\input{new_tables/tab_drop_2024}

\input{new_tables/tab_hpi}

\input{new_tables/tab_alt_functions}

\input{new_tables/tab_ipeds}

\end{document}

%% file: new_figures/fig_h1b_time.tex
\begin{figure}[htbp]
    \centering
    \caption{\textbf{H-1B Labor Condition Applications, 2017--2024}}
    \label{fig:h1b_time}
    \includegraphics[width=0.95\textwidth]{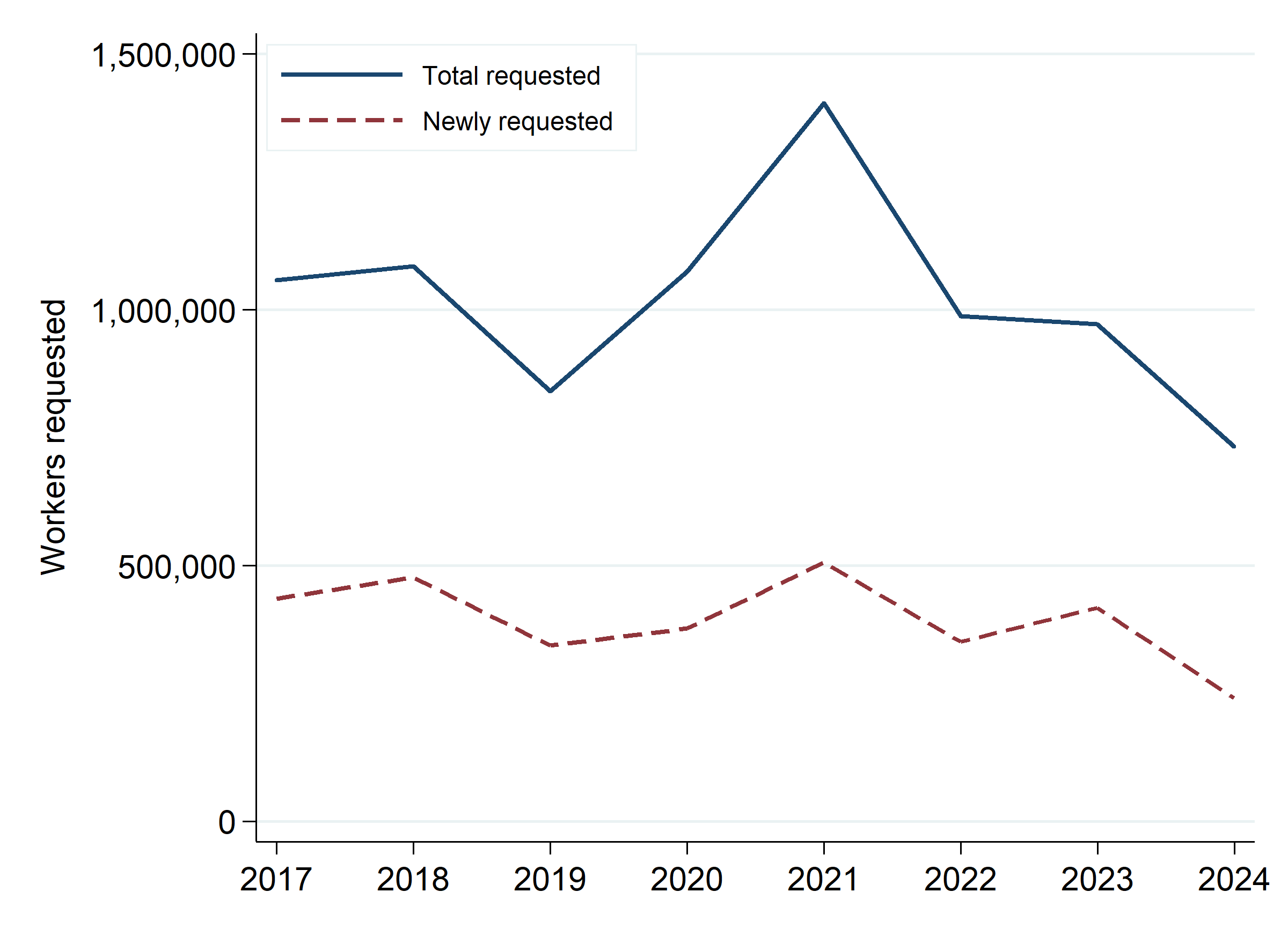}

    \vspace{0.5em}
    \begin{minipage}{\textwidth}
    \footnotesize Figure shows the total number of certified H-1B Labor Condition Application workers by year.
    \end{minipage}
\end{figure}

%% file: new_figures/fig_mortgage_rate.tex
\begin{figure}[!htbp]\centering
    \caption{\textbf{Mortgage-rate cycle and the national shock}}
    \label{fig:mortgage_rates}
    \includegraphics[width=\textwidth]{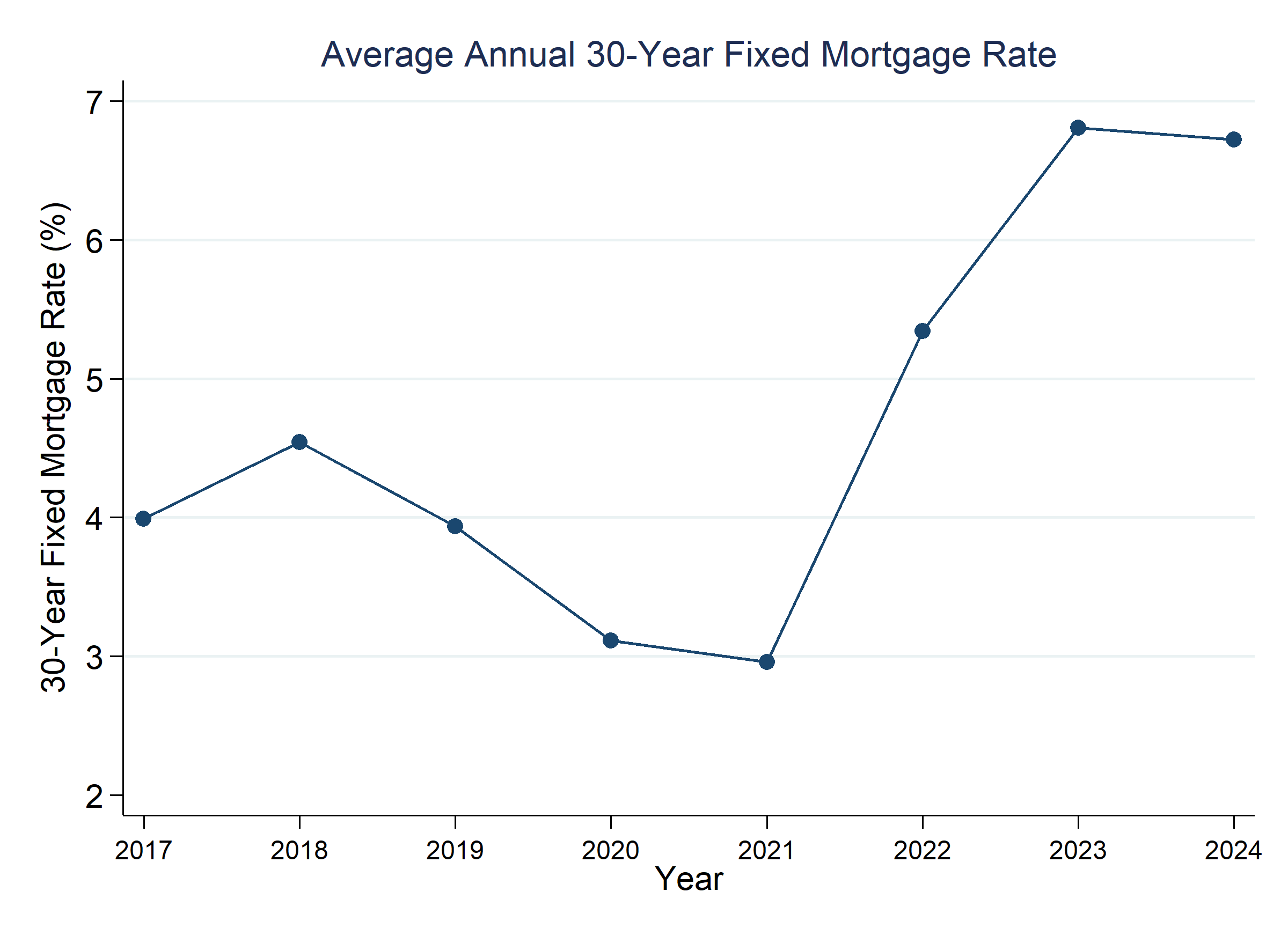}
    \begin{minipage}{\linewidth}
        \vspace{0.5em}
        \footnotesize This figure depicts the annual average 30-year fixed mortgage rate from 2017 to 2024, obtained from the Freddie Mac Primary Mortgage Market Survey via FRED (series \texttt{MORTGAGE30US}).
    \end{minipage}
\end{figure}

%% file: new_figures/fig_inmigration_cost.tex
\begin{figure}[htbp]
    \centering
    \caption{\textbf{In-Migration Mortgage Lock-In Costs by Commuting Zone}}
    \label{fig:inmigration_cost}
    \includegraphics[width=0.95\textwidth]{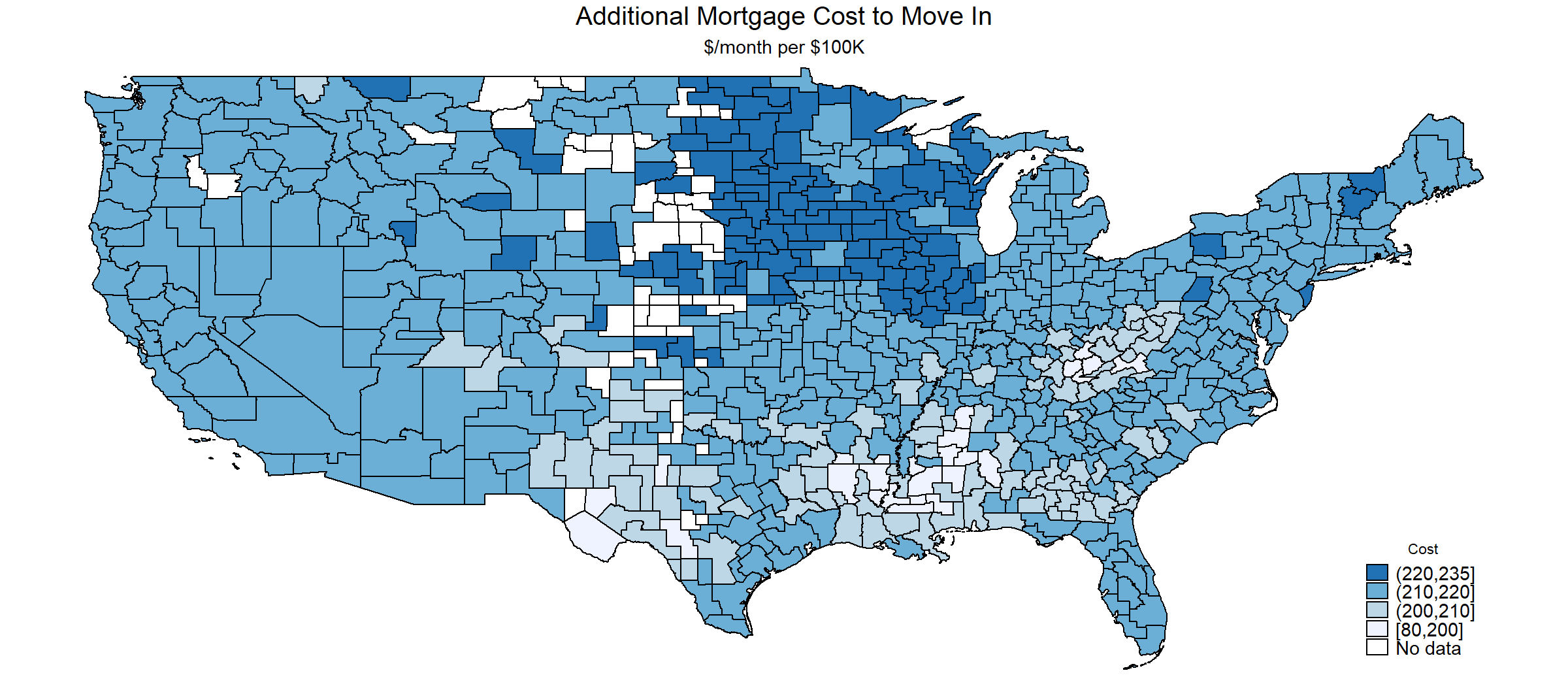}

    \vspace{0.5em}
    \begin{minipage}{\textwidth}
    \footnotesize Figure maps the additional monthly mortgage cost per \$100{,}000 of principal faced by workers moving into each commuting zone, weighted by historical migration patterns from origin commuting zones. White areas indicate missing data from IRS suppression of small migration flows.
    \end{minipage}
\end{figure}

%% file: new_figures/fig_map_costin_components.tex
\begin{figure}[htbp]
\centering
\caption{\textbf{Components of the In-Migration Mortgage-Payment Wedge}}
\label{fig:map_costin_components}
\begin{subfigure}{\linewidth}
\centering
\caption{Destination 2024 Mortgage Payment, $P^{new}_{d}$}
\label{fig:map_payment_new}
\includegraphics[width=\textwidth]{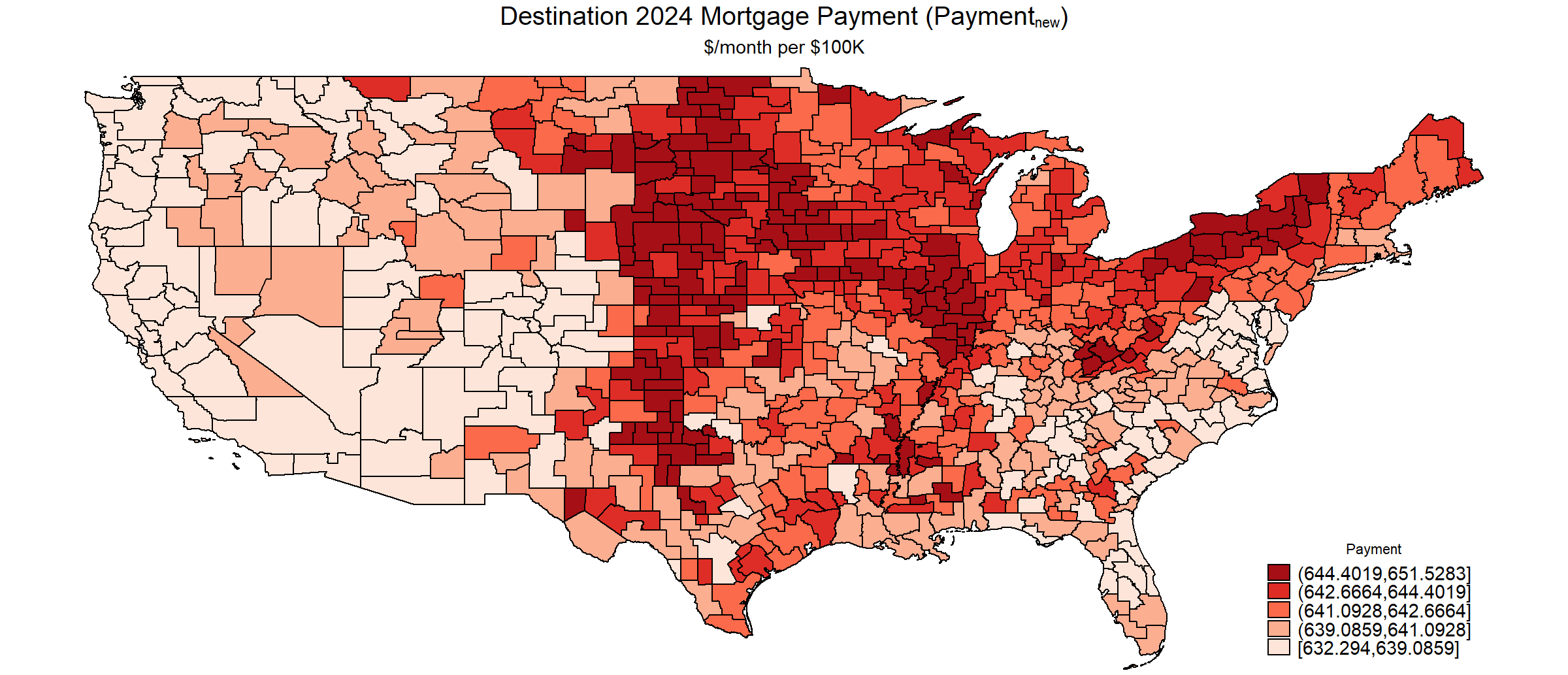}
\end{subfigure}
\vspace{1em}
\begin{subfigure}{\linewidth}
\centering
\caption{Locked-In Origin Payment, $WOP_{d}$}
\label{fig:map_wop}
\includegraphics[width=\textwidth]{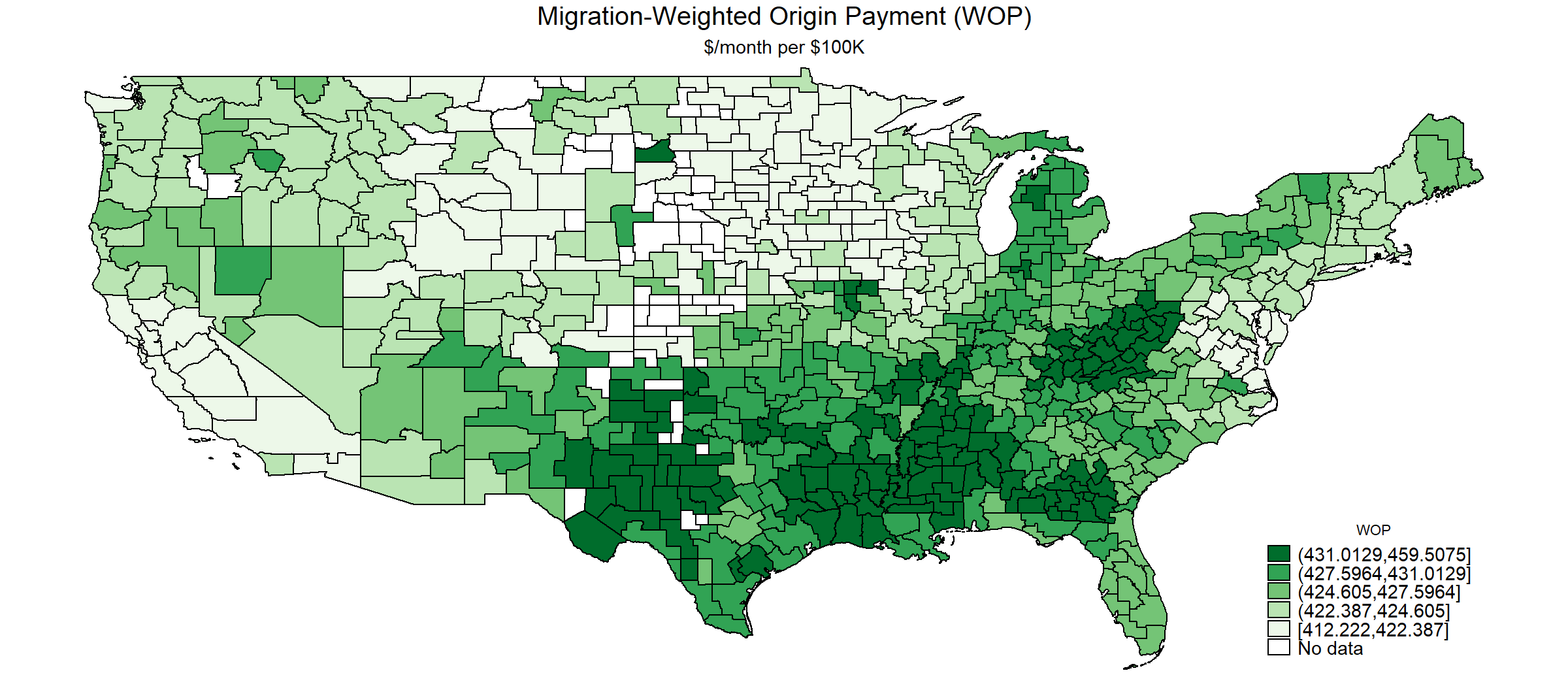}
\end{subfigure}
\begin{minipage}{\linewidth}
\vspace{0.5em}
\footnotesize This figure maps the two components of the in-migration mortgage-payment wedge, which decomposes as MPW$^{in}_{d}=P^{new}_{d}-WOP_{d}$ in equation~\eqref{eq:mpw_wop}. Panel (a) maps the destination component $P^{new}_{d}$, the counterfactual 2024-rate payment in commuting zone $d$, and panel (b) maps the origin component $WOP_{d}$, the migration-weighted actual 2020 to 2021 payment in $d$'s feeder commuting zones. Both are in dollars per month per \$100{,}000 of principal, and white areas denote commuting zones with missing wedge values.
\end{minipage}
\end{figure}

%% file: new_figures/fig_event_migration.tex
\begin{figure}[htbp]
\centering
\caption{\textbf{Event Study: Effect of the In-Migration Mortgage-Payment Wedge on College-Educated In-Migration}}
\label{fig:event_college_migration}
\includegraphics[width=\textwidth]{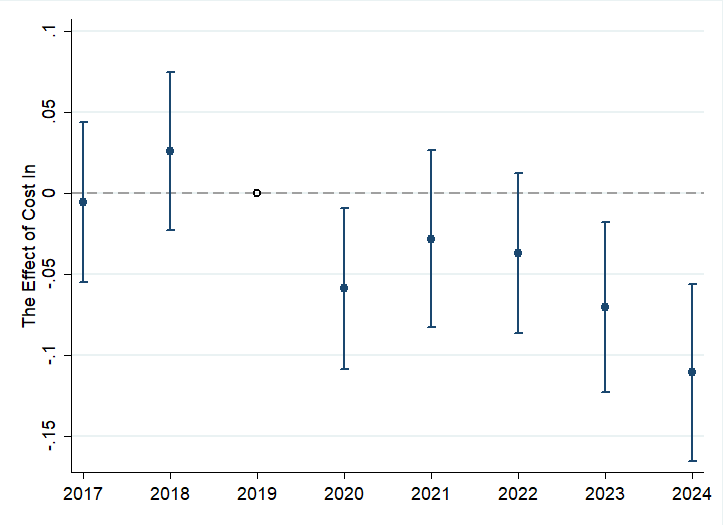}
\begin{minipage}{\linewidth}
\vspace{0.5em}
\footnotesize This figure plots event study coefficients from interactions of the in-migration mortgage-payment wedge with year indicators, with college-educated in-migration per 1,000 population as the outcome and 2019 as the reference year. Bars are 95\% confidence intervals. Fixed effects, controls, and clustering follow Table~\ref{tab:migration_costin}.
\end{minipage}
\end{figure}

%% file: new_figures/fig_event_h1b.tex
\begin{figure}[htbp]
\centering
\caption{\textbf{Event Study: Effect of the In-Migration Mortgage-Payment Wedge on Newly Requested H-1B Workers}}
\label{fig:event_h1b_workers_new_pe}
\includegraphics[width=\textwidth]{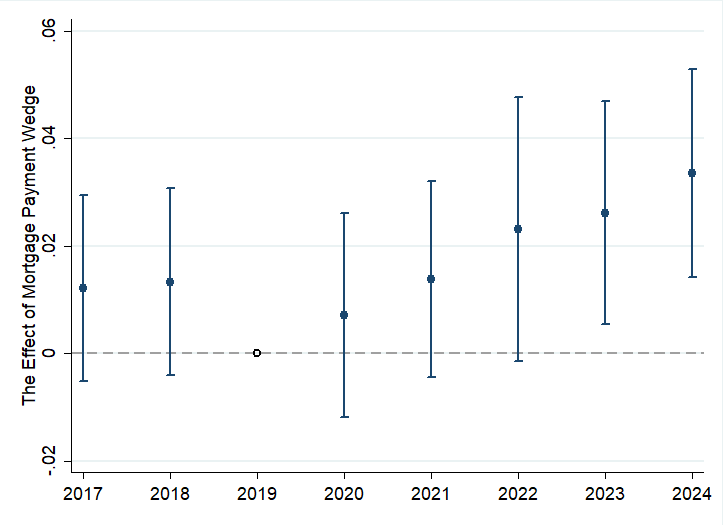}
\begin{minipage}{\linewidth}
\vspace{0.5em}
\footnotesize This figure plots event study coefficients from interactions of the in-migration mortgage-payment wedge with year indicators, with newly requested H-1B workers per 1,000 employed workers as the outcome and 2019 as the reference year. Bars are 95\% confidence intervals. Fixed effects, controls, and clustering follow Table~\ref{tab:migration_costin}.
\end{minipage}
\end{figure}

%% file: new_tables/tab_sumstats.tex
\begin{landscape}
\begin{table}[htbp]
\centering
\caption{\textbf{Summary Statistics}}
\label{tab:sumstats}
\begin{tabular}{p{7cm}p{12cm}cc}
\toprule
Variable & Description & Mean & Std. Dev. \\
\midrule
\multicolumn{4}{l}{\textit{Panel A: Mortgage Payment Wedge and Components (\$/month per \$100K)}} \\[0.5em]
Mortgage Payment Wedge ($MPW^{in}$) & Migration-weighted monthly mortgage-payment increase facing domestic movers into destination & 214.31 & 7.02 \\
2024 Mortgage Payment ($P^{new}$) & Counterfactual destination payment at 2024 rates & 641.42 & 3.00 \\
Locked-In Origin Payment ($WOP$) & Weighted average actual 2020--2021 payment in feeder origins & 427.11 & 6.85 \\
\midrule
\multicolumn{4}{l}{\textit{Panel B: H-1B Outcomes from LCAs (per 1,000 Employed)}} \\[0.5em]
Newly Requested Workers & Certified positions in the new-employment category & 0.636 & 2.028 \\
Total Requested Workers & Certified positions across all six employment-action categories & 1.565 & 5.199 \\
New LCAs & Certified applications with at least one new-employment position & 0.391 & 0.811 \\
New Filing Firms & Distinct firms with at least one certified new-employment LCA & 0.139 & 0.153 \\
\midrule
\multicolumn{4}{l}{\textit{Panel C: H-1B Outcomes from USCIS (per 1,000 Employed)}} \\[0.5em]
Newly Approved Workers & USCIS approvals in the new-employment category & 0.277 & 0.882 \\
Total Approved Workers & USCIS approvals across all employment-action categories & 0.784 & 2.793 \\
New Petitions & USCIS new-employment approvals plus new-employment denials & 0.301 & 0.941 \\
Filing Firms (USCIS) & Distinct employers with any adjudicated H-1B petition & 0.215 & 0.339 \\
\midrule
\multicolumn{4}{l}{\textit{Panel D: Migration Outcomes (per 1,000 Population)}} \\[0.5em]
Total Domestic In-Migration & ACS domestic in-migrants & 82.851 & 31.903 \\
College In-Migration & ACS college-educated domestic in-migrants & 14.727 & 7.015 \\
Non-College In-Migration & ACS non-college domestic in-migrants & 68.124 & 28.693 \\
College Owner In-Migration & ACS college-educated owner in-migrants & 7.607 & 4.306 \\
College Renter In-Migration & ACS college-educated renter in-migrants & 7.120 & 4.282 \\
\midrule
\multicolumn{4}{r}{\textit{(Continued on next page)}} \\
\end{tabular}
\end{table}
\clearpage
\begin{table}[htbp]
\ContinuedFloat
\centering
\caption{\textbf{Summary Statistics} (continued)}
\begin{tabular}{p{7cm}p{12cm}cc}
\toprule
Variable & Description & Mean & Std. Dev. \\
\midrule
\multicolumn{4}{l}{\textit{Panel D: Migration Outcomes (per 1,000 Population), continued}} \\[0.5em]
College Inflows from Abroad & ACS college-educated foreign inflows & 0.979 & 1.242 \\
Non-College Inflows from Abroad & ACS non-college foreign inflows & 2.877 & 2.602 \\
\midrule
\multicolumn{4}{l}{\textit{Panel E: Commuting-Zone Controls (Shares unless noted)}} \\[0.5em]
Age 18--34 & Share of CZ population age 18--34 & 0.217 & 0.028 \\
Age 35--54 & Share age 35--54 & 0.240 & 0.015 \\
Age 55+ & Share age 55+ & 0.320 & 0.045 \\
Black & Share Black & 0.084 & 0.116 \\
Asian & Share Asian & 0.021 & 0.038 \\
Hispanic & Share Hispanic & 0.120 & 0.146 \\
Foreign-Born & Share foreign-born & 0.069 & 0.058 \\
College-Plus & Share with bachelor's degree or higher & 0.186 & 0.058 \\
STEM Degree & Share with STEM bachelor's degree & 0.177 & 0.042 \\
Teleworkable & Employment-weighted teleworkable share & 0.344 & 0.049 \\
Homeownership Rate & Share of households owning their home & 0.648 & 0.061 \\
CZ Unemployment Rate (\%) & BLS LAUS unemployment rate, in percent & 4.476 & 1.786 \\
Baseline Tech Employment Share & 2017--2019 CZ share in seven 4-digit NAICS tech industries & 0.017 & 0.016 \\
\bottomrule
\end{tabular}
\vspace{0.5em}
\begin{minipage}{\linewidth}
\footnotesize Summary statistics for the estimation sample at the commuting-zone-year level over 2017--2024. Panels A, B, D, and E use 5,475 observations. Panel C uses the 4,899 cells with non-missing USCIS outcomes, and the corresponding regression sample further excludes 11 cells with missing controls (N = 4,888).
\end{minipage}
\end{table}
\end{landscape}

%% file: new_tables/tab_migration.tex
\begin{landscape}
\begin{table}[htbp]
\centering
\caption{\textbf{Migration Response to the In-Migration Mortgage-Payment Wedge}}
\label{tab:migration_costin}
\begin{tabular}{lcccccc}
\toprule
\multicolumn{7}{c}{\textit{Panel A}} \\
\cmidrule(lr){2-6}
& (1) & (2) & (3) & (4) & (5) & \\
& Total        & Domestic & Domestic    & Domestic & Domestic & \\
& In-Migration & College  & Non-College & Owner    & Renter   & \\
&              &          &             & College  & College  & \\
\midrule
Mortgage Payment Wedge $\times$ Post & $-$0.117** & $-$0.059*** & $-$0.058 & $-$0.059*** & $-$0.001 & \\
                                     & (0.053)    & (0.015)     & (0.049)  & (0.010)     & (0.009)  & \\
\midrule
CZ and Year Fixed Effects & Yes   & Yes   & Yes   & Yes   & Yes   & \\
Time-Varying Controls     & Yes   & Yes   & Yes   & Yes   & Yes   & \\
Observations              & 5,475 & 5,475 & 5,475 & 5,475 & 5,475 & \\
\midrule
\multicolumn{7}{c}{\textit{Panel B}} \\
\cmidrule(lr){2-7}
& (1) & (2) & (3) & (4) & (5) & (6) \\
& From    & From        & From    & From        & From        & From        \\
& Abroad  & Abroad      & Abroad  & Abroad      & Abroad      & Abroad      \\
& College & Non-College & College & College     & Non-College & Non-College \\
&         &             & Citizen & Non-Citizen & Citizen     & Non-Citizen \\
\midrule
Mortgage Payment Wedge $\times$ Post & 0.015*** & $-$0.003 & 0.004   & 0.013*** & 0.005   & $-$0.012* \\
                                     & (0.004)  & (0.009)  & (0.003) & (0.004)  & (0.006) & (0.007)   \\
\midrule
CZ and Year Fixed Effects & Yes   & Yes   & Yes   & Yes   & Yes   & Yes   \\
Time-Varying Controls     & Yes   & Yes   & Yes   & Yes   & Yes   & Yes   \\
Observations              & 5,475 & 5,475 & 5,475 & 5,475 & 5,475 & 5,475 \\
\bottomrule
\end{tabular}
\vspace{0.5em}
\begin{minipage}{\linewidth}
\footnotesize Estimates of the effect of the in-migration mortgage-payment wedge on in-migration per 1,000 population, with column subgroups given in the headers. The Mortgage Payment Wedge, MPW$^{in}_{d}=P^{new}_{d}-WOP_{d}$, is the migration-weighted monthly mortgage-payment increase faced by domestic movers into destination $d$, in dollars per month per \$100{,}000 of principal. Post indicates years 2022 to 2024. All specifications include commuting zone and year fixed effects and time-varying controls (age shares, race/ethnicity shares, foreign-born share, college share, STEM share, teleworkable share, homeownership rate, and CZ unemployment rate). Standard errors clustered at the commuting-zone level are in parentheses. *** p$<$0.01, ** p$<$0.05, * p$<$0.10.
\end{minipage}
\end{table}
\end{landscape}

%% file: new_tables/tab_mig_decomp.tex
\begin{landscape}
\begin{table}[htbp]
\centering
\caption{\textbf{Decomposition of the In-Migration Mortgage-Payment Wedge}}
\label{tab:migration_decomposition}
\begin{tabular}{lccccccc}
\toprule
& (1) & (2) & (3) & (4) & (5) & (6) & (7) \\
& \multicolumn{7}{c}{\textit{In-Migration per 1,000 Population}} \\
\cmidrule(lr){2-8}
& Total & College & Non-College & Owner   & Renter  & From    & From        \\
&       &         &             & College & College & Abroad  & Abroad      \\
&       &         &             &         &         & College & Non-College \\
\midrule
\multicolumn{8}{l}{\textit{2024 Mortgage Payment vs.\ Locked-In Origin Payment}} \\
\midrule
2024 Mortgage Payment $\times$ Post     & $-$0.351** & $-$0.025 & $-$0.326*** & $-$0.060** & 0.034     & 0.028***   & 0.021    \\
                                        & (0.135)    & (0.034)  & (0.122)     & (0.023)    & (0.022)   & (0.010)    & (0.021)  \\
[0.5em]
Locked-In Origin Payment $\times$ Post  & 0.085      & 0.064*** & 0.021       & 0.058***   & 0.006     & $-$0.014*** & 0.006   \\
                                        & (0.055)    & (0.015)  & (0.051)     & (0.010)    & (0.010)   & (0.004)    & (0.009)  \\
\midrule
Observations                            & 5,475 & 5,475 & 5,475 & 5,475 & 5,475 & 5,475 & 5,475 \\
\cmidrule{1-8}
CZ and Year Fixed Effects   & Yes & Yes & Yes & Yes & Yes & Yes & Yes \\
Time-Varying Controls       & Yes & Yes & Yes & Yes & Yes & Yes & Yes \\
\bottomrule
\end{tabular}
\vspace{0.5em}
\begin{minipage}{\linewidth}
\footnotesize This table replicates Table~\ref{tab:migration_costin} with $P^{new}_{d}$ and $WOP_{d}$ entered jointly in place of $MPW^{in}_{d}$, where $MPW^{in}_{d}=P^{new}_{d}-WOP_{d}$. $P^{new}_{d}$ is the destination's counterfactual monthly payment at 2024 rates. $WOP_{d}$ is the migration-weighted actual 2020 to 2021 payment in destination $d$'s historical origins. Both are measured in dollars per month per \$100{,}000 of principal. The rest of the specification follows from Table~\ref{tab:migration_costin}. *** p$<$0.01, ** p$<$0.05, * p$<$0.10.
\end{minipage}
\end{table}
\end{landscape}

%% file: new_tables/tab_h1b_cz.tex
\begin{table}[htbp]
\centering
\caption{\textbf{H-1B Activity Response to the In-Migration Mortgage-Payment Wedge}}
\label{tab:h1b_cz}
\begin{tabular}{lcccc}
\toprule
& (1)       & (2)       & (3)  & (4)     \\
& Newly     & Total     & New  & New     \\
& Requested & Requested & LCAs & Filing  \\
& Workers   & Workers   &      & Firms   \\
\midrule
Mortgage Payment Wedge $\times$ Post & 0.018*** & 0.025*** & 0.011*** & 0.0021*** \\
                                     & (0.006)  & (0.009)  & (0.004)  & (0.0005)  \\
[0.5em]
Observations               & 5,475 & 5,475 & 5,475 & 5,475 \\
\cmidrule{1-5}
CZ and Year Fixed Effects  & Yes & Yes & Yes & Yes \\
Time-Varying Controls      & Yes & Yes & Yes & Yes \\
\bottomrule
\end{tabular}
\vspace{0.5em}
\begin{minipage}{\linewidth}
\footnotesize This table replicates Table~\ref{tab:migration_costin} with H-1B activity outcomes in place of the migration outcomes, scaled per 1,000 employed workers in the commuting zone. Newly Requested Workers (col.\ 1) counts initial new-employment H-1B requests. Total Requested Workers (col.\ 2) counts all H-1B labor requests, including continuations. New LCAs (col.\ 3) counts new Labor Condition Applications, the first step of the H-1B process. New Filing Firms (col.\ 4) counts distinct firms filing certified new-employment LCAs in the commuting zone and year. The rest of the specification follows from Table~\ref{tab:migration_costin}. *** p$<$0.01, ** p$<$0.05, * p$<$0.10.
\end{minipage}
\end{table}

%% file: new_tables/tab_h1b_cz_decomposition.tex
\begin{table}[htbp]
\centering
\caption{\textbf{Decomposition of the H-1B Response to the In-Migration Mortgage-Payment Wedge}}
\label{tab:h1b_decomposition}
\begin{tabular}{lcccc}
\toprule
& (1)       & (2)       & (3)  & (4)     \\
& Newly     & Total     & New  & New     \\
& Requested & Requested & LCAs & Filing  \\
& Workers   & Workers   &      & Firms   \\
\midrule
2024 Mortgage Payment $\times$ Post     & 0.036***    & 0.098***    & 0.009      & 0.0040***  \\
                                        & (0.012)     & (0.023)     & (0.006)    & (0.0013)   \\
[0.5em]
Locked-In Origin Payment $\times$ Post  & $-$0.016**  & $-$0.015    & $-$0.011** & $-$0.0018*** \\
                                        & (0.006)     & (0.010)     & (0.005)    & (0.0006)     \\
\midrule
Observations               & 5,475 & 5,475 & 5,475 & 5,475 \\
\cmidrule{1-5}
CZ and Year Fixed Effects  & Yes & Yes & Yes & Yes \\
Time-Varying Controls      & Yes & Yes & Yes & Yes \\
\bottomrule
\end{tabular}
\vspace{0.5em}
\begin{minipage}{\linewidth}
\footnotesize This table replicates Table~\ref{tab:h1b_cz} with $P^{new}_{d}$ and $WOP_{d}$ entered jointly in place of $MPW^{in}_{d}$, where $MPW^{in}_{d}=P^{new}_{d}-WOP_{d}$. $P^{new}_{d}$ is the destination's counterfactual monthly payment at 2024 rates. $WOP_{d}$ is the migration-weighted actual 2020 to 2021 payment in destination $d$'s historical origins. Both are measured in dollars per month per \$100{,}000 of principal. The rest of the specification follows from Table~\ref{tab:h1b_cz}. *** p$<$0.01, ** p$<$0.05, * p$<$0.10.
\end{minipage}
\end{table}

%% file: new_tables/tab_cz_soc_h1b.tex
\begin{table}[htbp]
\centering
\caption{\textbf{H-1B Activity Response to the In-Migration Mortgage-Payment Wedge, by SOC Demand Shock}}
\label{tab:h1b_cz_soc_bartik}
\begin{tabular}{lcccc}
\toprule
& (1)       & (2)       & (3)  & (4)     \\
& Newly     & Total     & New  & New     \\
& Requested & Requested & LCAs & Filing  \\
& Workers   & Workers   &      & Firms   \\
\midrule
Mortgage Payment Wedge $\times$ Post $\times$ Bartik & 0.048*** & 0.106*** & 0.021*** & 0.0014    \\
                                                     & (0.014)  & (0.020)  & (0.005)  & (0.0017)  \\
[0.5em]
Observations               & 43,889 & 43,889 & 43,889 & 43,889 \\
\cmidrule{1-5}
CZ $\times$ Year FE        & Yes & Yes & Yes & Yes \\
CZ $\times$ SOC FE         & Yes & Yes & Yes & Yes \\
SOC $\times$ Year FE       & Yes & Yes & Yes & Yes \\
\bottomrule
\end{tabular}
\vspace{0.5em}
\begin{minipage}{\linewidth}
\footnotesize This table reports the differential effect of the wedge on H-1B activity in commuting zones exposed to stronger SOC-level demand shocks. The unit of observation is the commuting-zone-by-SOC-by-year cell. Bartik is the demeaned SOC-specific shift-share demand shock in percentage points, built from national SOC employment growth and each commuting zone's pre-period SOC mix. The reported row is the triple interaction $MPW^{in}_c \times Post_t \times \widetilde{B}_{cst}$. Fixed effects are commuting-zone $\times$ year, commuting-zone $\times$ SOC, and SOC $\times$ year, with all lower-order terms absorbed. The rest of the specification follows from Tables~\ref{tab:migration_costin} and~\ref{tab:h1b_cz}. *** p$<$0.01, ** p$<$0.05, * p$<$0.10.
\end{minipage}
\end{table}

%% file: new_tables/tab_uscis.tex
\begin{table}[htbp]
\centering
\caption{\textbf{USCIS H-1B Petition and Approval Response to the In-Migration Mortgage-Payment Wedge}}
\label{tab:uscis}
\begin{tabular}{lcccc}
\toprule
& (1)      & (2)      & (3)       & (4)    \\
& Newly    & Total    & New       & Filing \\
& Approved & Approved & Petitions & Firms  \\
& Workers  & Workers  &           &        \\
\midrule
Mortgage Payment Wedge $\times$ Post & 0.0043* & 0.0078** & 0.0049* & 0.0027*** \\
                                     & (0.0024) & (0.0031) & (0.0025) & (0.0005)  \\
[0.5em]
Observations               & 4,888 & 4,888 & 4,888 & 4,888 \\
\cmidrule{1-5}
CZ and Year Fixed Effects  & Yes & Yes & Yes & Yes \\
Time-Varying Controls      & Yes & Yes & Yes & Yes \\
\bottomrule
\end{tabular}
\vspace{0.5em}
\begin{minipage}{\linewidth}
\footnotesize This table replicates Table~\ref{tab:h1b_cz} with the four USCIS petition and approval outcomes in place of the LCA outcomes, scaled per 1,000 employed workers in the commuting zone. Newly Approved Workers (col.\ 1) is new-employment approvals. Total Approved Workers (col.\ 2) is total approvals across all employment-action categories. New Petitions (col.\ 3) is new-employment approvals plus new-employment denials. Filing Firms (col.\ 4) is the number of distinct employers filing H-1B petitions. The estimation sample is restricted to commuting zones with non-missing USCIS and LCA outcomes. The rest of the specification follows from Table~\ref{tab:migration_costin}. *** p$<$0.01, ** p$<$0.05, * p$<$0.10.
\end{minipage}
\end{table}

%% file: new_tables/tab_iv_wop.tex
\begin{table}[htbp]
\centering
\caption{\textbf{Instrumenting the Locked-In Origin Payment Component of the Wedge}}
\label{tab:iv_wop}
\begin{tabular}{lccc}
\toprule
\multicolumn{4}{c}{\textit{Panel A: Migration Outcomes (per 1,000 population)}} \\
\cmidrule(lr){2-4}
& (1) & (2) & (3) \\
& College      & College & College \\
& In-Migration & Owners  & Renters \\
\midrule
2024 Mortgage Payment $\times$ Post    & $-$0.032 & $-$0.068*** & 0.036* \\
                                       & (0.034)  & (0.023)  & (0.022)  \\
Locked-In Origin Payment $\times$ Post & 0.082*** & 0.082*** & 0.001    \\
                                       & (0.025)  & (0.018)  & (0.015)  \\
\midrule
CZ and Year Fixed Effects   & Yes   & Yes   & Yes   \\
Time-Varying Controls       & Yes   & Yes   & Yes   \\
Observations                & 5,472 & 5,472 & 5,472 \\
\midrule
\multicolumn{4}{c}{\textit{Panel B: H-1B Outcomes (per 1,000 employed)}} \\
\cmidrule(lr){2-4}
& (1) & (2) & (3) \\
& Newly Requested & New  & New Filing \\
& Workers         & LCAs & Firms      \\
\midrule
2024 Mortgage Payment $\times$ Post    & 0.037***    & 0.009       & 0.0046***    \\
                                       & (0.012)     & (0.006)     & (0.0013)     \\
Locked-In Origin Payment $\times$ Post & $-$0.018*** & $-$0.011*** & $-$0.0035*** \\
                                       & (0.007)     & (0.004)     & (0.0008)     \\
\midrule
CZ and Year Fixed Effects   & Yes   & Yes   & Yes   \\
Time-Varying Controls       & Yes   & Yes   & Yes   \\
Observations                & 5,472 & 5,472 & 5,472 \\
\bottomrule
\end{tabular}
\vspace{0.5em}
\begin{minipage}{\linewidth}
\footnotesize Two-stage least squares estimates on the commuting-zone-year panel. The Locked-In Origin Payment $\times$ Post (endogenous regressor) and the 2024 Mortgage Payment $\times$ Post (included instrument), the two components of $MPW^{in}_c$ in equation~\eqref{eq:mpw_wop}, are entered jointly. The excluded instrument is the gravity-and-lender-predicted Locked-In Origin Payment described in Section~\ref{sec:iv}. The first-stage coefficient is 49.63 and the cluster-robust Kleibergen-Paap first-stage $F$ is 392.16, on 5,472 commuting-zone-year cells across 684 commuting zones. The rest of the specification follows from Tables~\ref{tab:migration_costin} and~\ref{tab:h1b_cz}. *** p$<$0.01, ** p$<$0.05, * p$<$0.10.
\end{minipage}
\end{table}

%% file: new_figures/fig_h1b_map.tex
\begin{figure}[htbp]
    \centering
    \caption{\textbf{Geographic Distribution of H-1B Workers per Capita}}
    \label{fig:h1b_map}
    \includegraphics[width=0.95\textwidth]{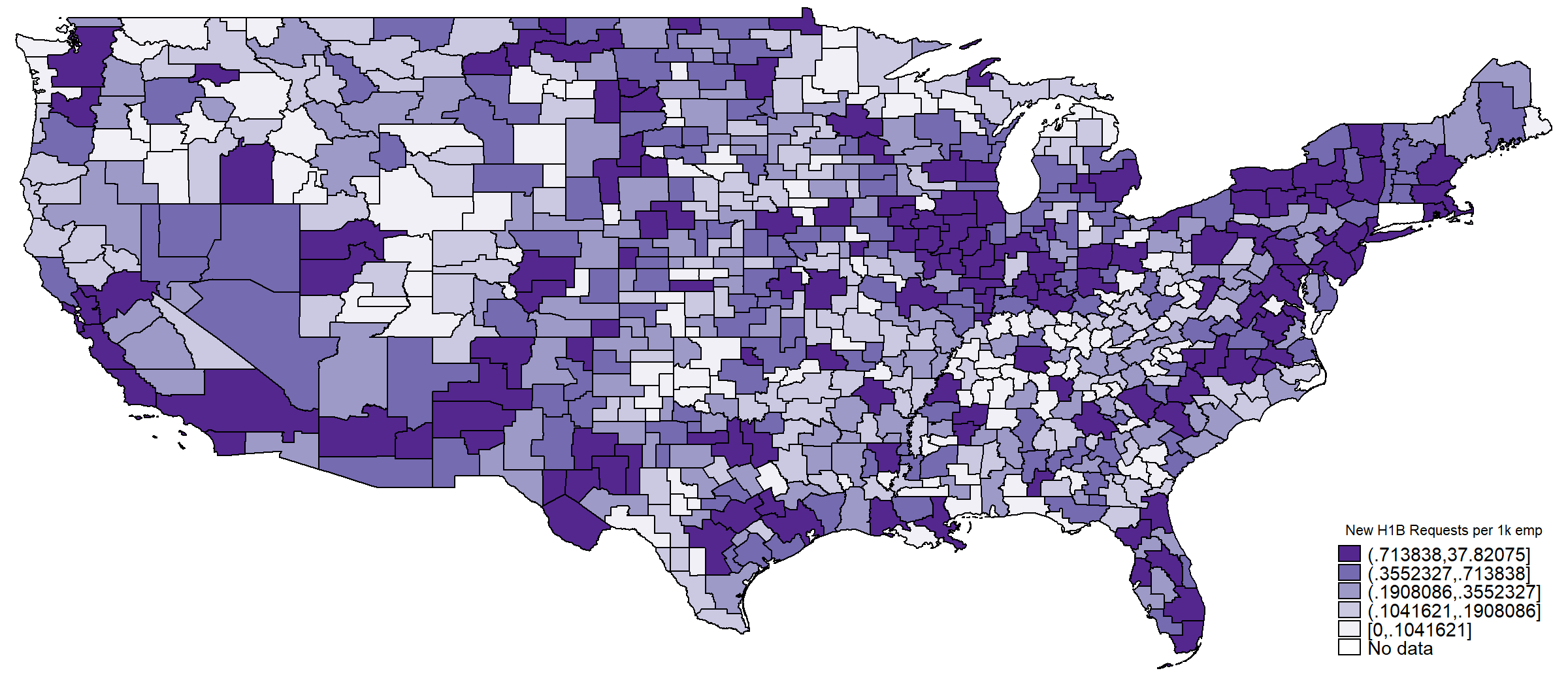}

    \vspace{0.5em}
    \begin{minipage}{\textwidth}
    \footnotesize This figure maps average H-1B workers per 1,000 population by commuting zone over 2017 to 2024, with commuting zones grouped into quintiles.
    \end{minipage}
\end{figure}

%% file: new_figures/fig_event_mig_components.tex
\begin{figure}[htbp]
\centering
\caption{\textbf{Event Studies for the Components of MPW$^{in}$}}
\label{fig:event_mig_components}
\begin{subfigure}{0.49\linewidth}
\centering
\caption{2024 Mortgage Payment, $P^{new}_{d}$}
\label{fig:event_payment_new}
\includegraphics[width=\linewidth]{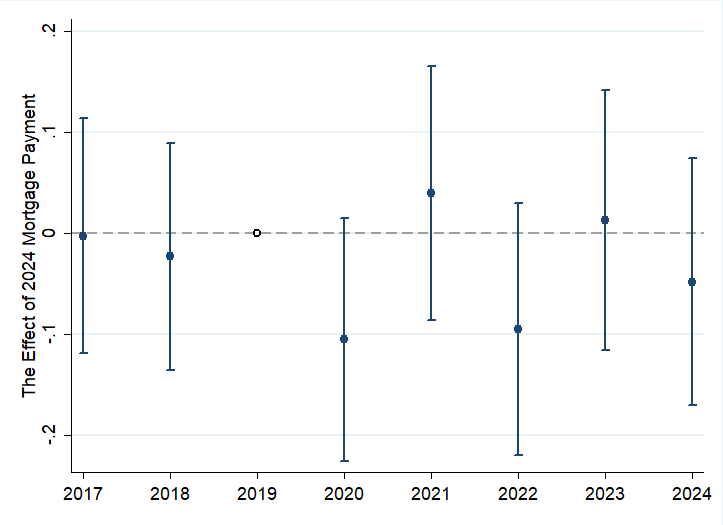}
\end{subfigure}
\hfill
\begin{subfigure}{0.49\linewidth}
\centering
\caption{Locked-In Origin Payment, $WOP_{d}$}
\label{fig:event_wop}
\includegraphics[width=\linewidth]{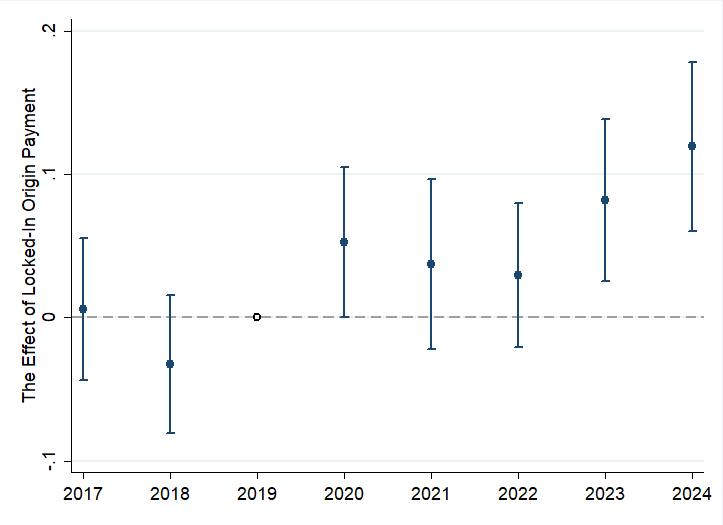}
\end{subfigure}
\begin{minipage}{\linewidth}
\vspace{0.5em}
\footnotesize Event studies of college-educated in-migration per 1,000 population on the two components of $MPW^{in}_d$, $P^{new}_d$ (panel a) and $WOP_d$ (panel b), each interacted with year indicators in a single regression. Reference year 2019; bars are 95\% confidence intervals; the rest of the specification follows from Table~\ref{tab:migration_costin}.
\end{minipage}
\end{figure}

%% file: new_figures/fig_event_h1b_cz_components.tex
\begin{figure}[htbp]
\centering
\caption{\textbf{Event Studies for the Components of MPW$^{in}$, Newly Requested H-1B Workers}}
\label{fig:event_h1b_components}
\begin{subfigure}{0.49\linewidth}
\centering
\caption{2024 Mortgage Payment, $P^{new}_{d}$}
\label{fig:event_h1b_payment_new}
\includegraphics[width=\linewidth]{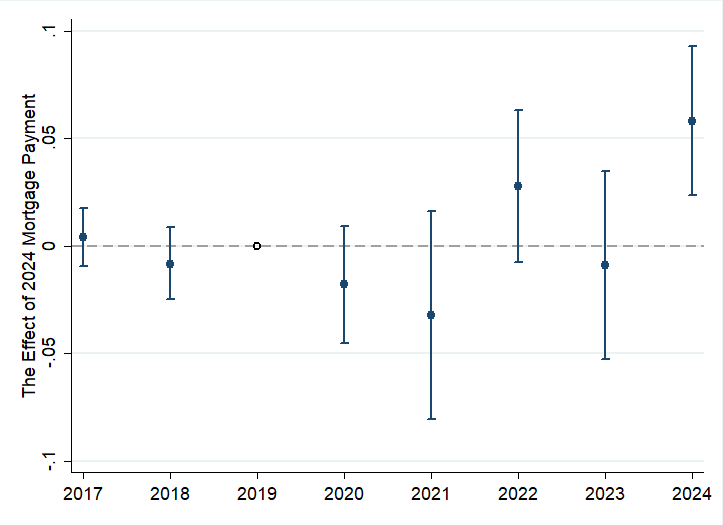}
\end{subfigure}
\hfill
\begin{subfigure}{0.49\linewidth}
\centering
\caption{Locked-In Origin Payment, $WOP_{d}$}
\label{fig:event_h1b_wop}
\includegraphics[width=\linewidth]{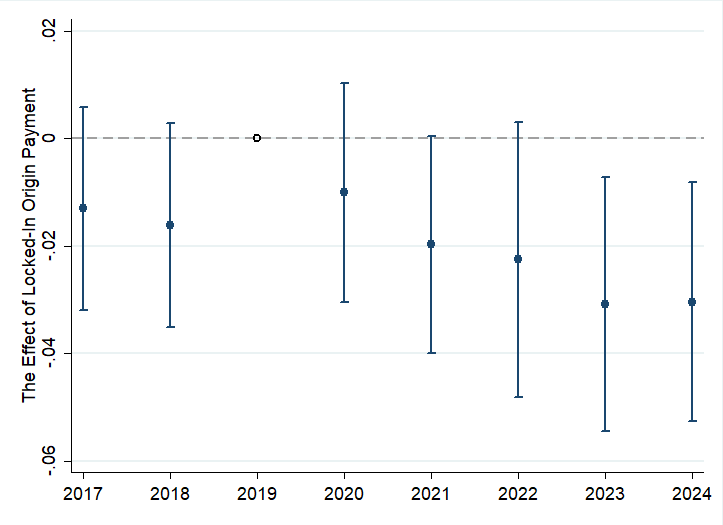}
\end{subfigure}
\begin{minipage}{\linewidth}
\vspace{0.5em}
\footnotesize Event studies of newly requested H-1B workers per 1,000 employed workers on the two components of $MPW^{in}_d$, $P^{new}_d$ (panel a) and $WOP_d$ (panel b), each interacted with year indicators in a single regression. Reference year 2019; bars are 95\% confidence intervals; the rest of the specification follows from Table~\ref{tab:h1b_cz}.
\end{minipage}
\end{figure}

%% file: new_figures/fig_loo_soc.tex
\begin{figure}[htbp]
\centering
\caption{\textbf{Leave-One-Out Robustness by SOC Major Group}}
\label{fig:leaveoneout_soc}
\includegraphics[width=\textwidth]{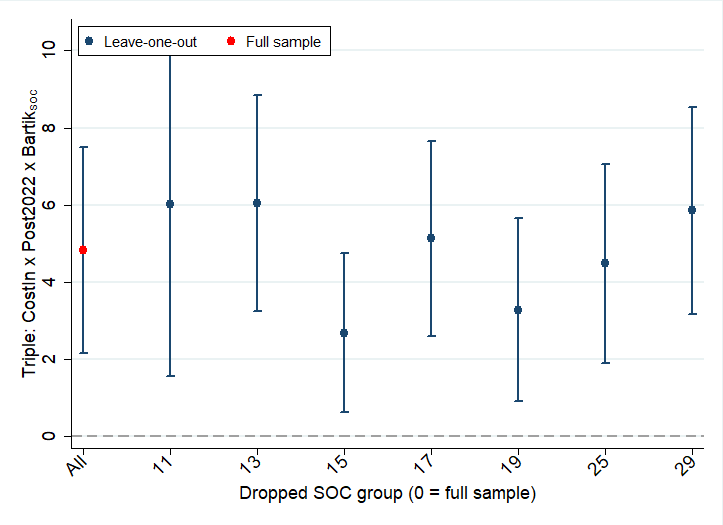}
\begin{minipage}{\linewidth}
\vspace{0.5em}
\footnotesize Leave-one-out version of Table~\ref{tab:h1b_cz_soc_bartik}: the coefficient on the Mortgage Payment Wedge $\times$ Post $\times$ Bartik triple interaction (outcome: newly requested H-1B workers per 1,000 employed workers), re-estimated while dropping one two-digit SOC major group at a time. The red marker labeled All is the full-sample estimate; each navy marker drops the SOC major group on the horizontal axis. Bars are 95\% confidence intervals; the rest of the specification follows from Table~\ref{tab:h1b_cz_soc_bartik}.
\end{minipage}
\end{figure}

%% file: new_figures/fig_wop_netperm.tex
\begin{figure}[htbp]
\centering
\caption{\textbf{Network-Permutation Placebo for the In-Migration Mortgage-Payment Wedge}}
\label{fig:robust_wop_netperm}
\begin{subfigure}{0.49\linewidth}
\centering
\caption{College In-Migration}
\label{fig:robust_wop_netperm_mig}
\includegraphics[width=\linewidth]{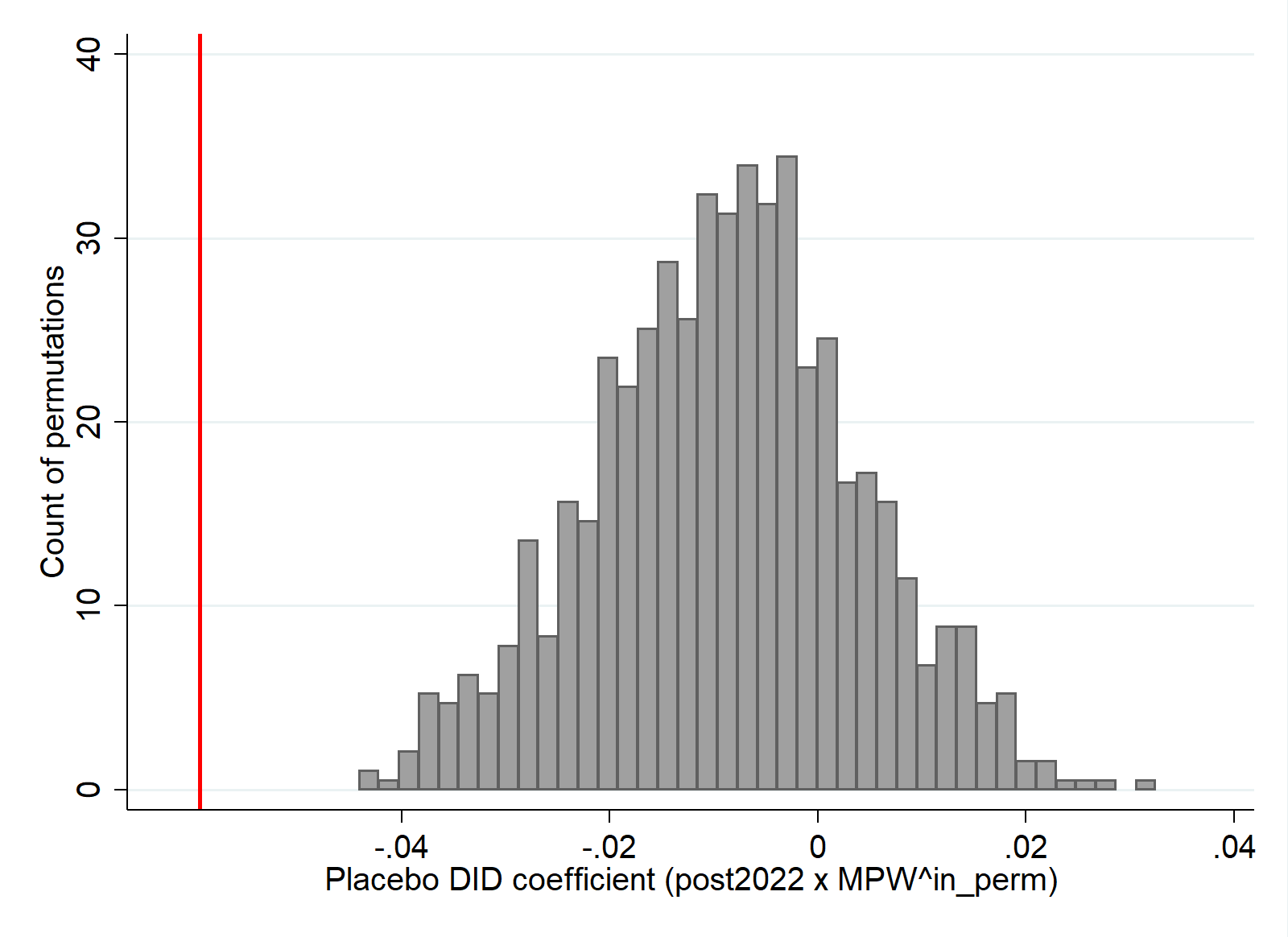}
\end{subfigure}
\hfill
\begin{subfigure}{0.49\linewidth}
\centering
\caption{Newly Requested H-1B Workers}
\label{fig:robust_wop_netperm_h1b}
\includegraphics[width=\linewidth]{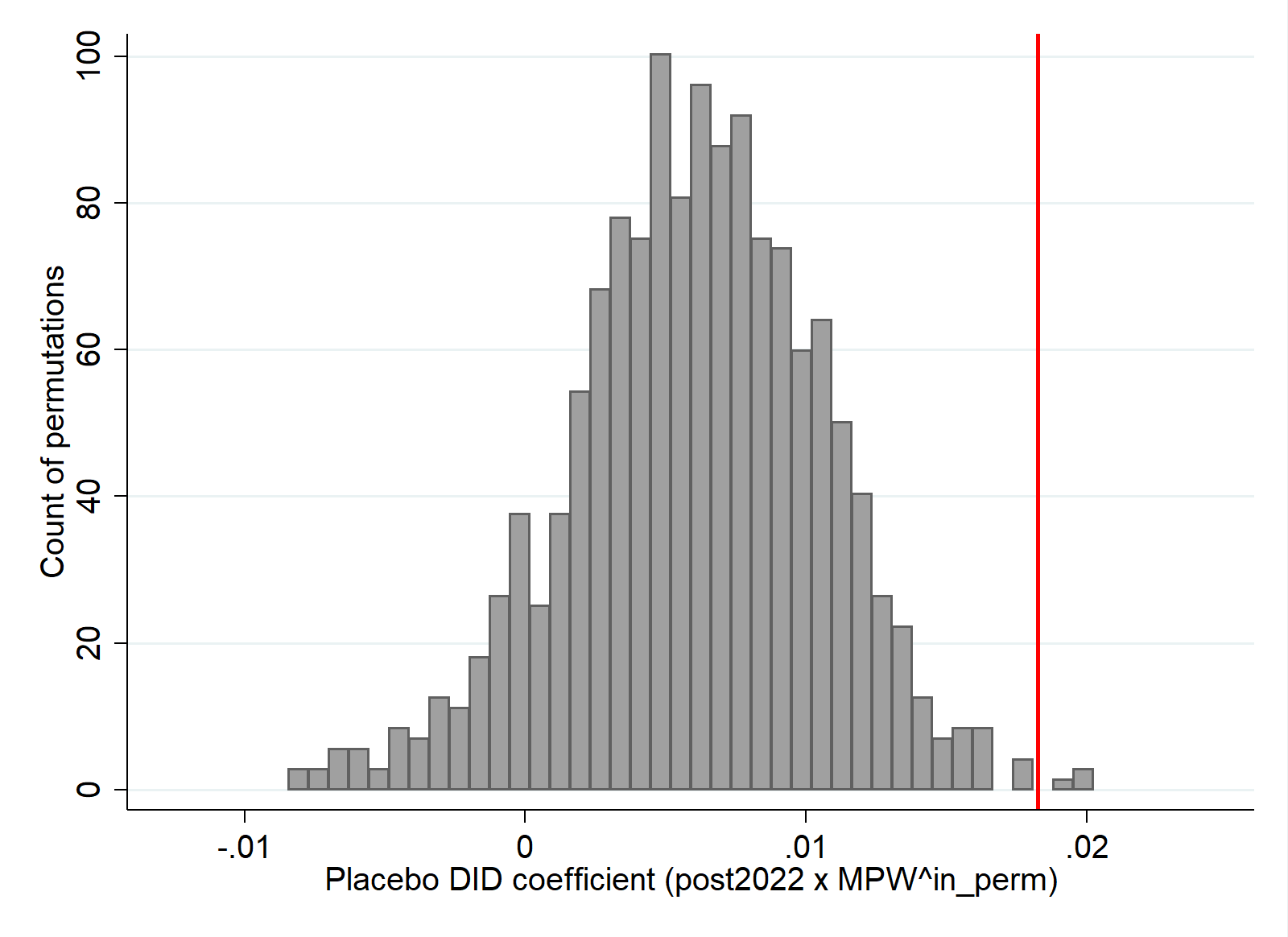}
\end{subfigure}
\begin{minipage}{\linewidth}
\vspace{0.5em}
\footnotesize Distribution of placebo difference-in-differences coefficients on Post $\times$ MPW$^{in,perm}_c$ across 1{,}000 permutations that reshuffle $WOP_d$ across destinations, holding $P^{new}_d$ fixed and recomputing $MPW^{in,perm}_d = P^{new}_d - WOP^{perm}_d$, which breaks the match between a destination and its historical feeder network. Outcomes: college in-migration (panel a) and newly requested H-1B workers (panel b). The vertical red line marks the baseline estimate ($-0.059$ in panel a, $0.018$ in panel b), with two-sided centered placebo $p$-values below 0.001 and 0.011 respectively. The rest of the specification follows from Tables~\ref{tab:migration_costin} and~\ref{tab:h1b_cz}.
\end{minipage}
\end{figure}

%% file: new_tables/tab_h1b_tech_intensity.tex
\begin{landscape}
\begin{table}[htbp]
\centering
\caption{\textbf{Heterogeneity by Local Tech Employment Intensity}}
\label{tab:h1b_heterogeneity_tech}
\setlength{\tabcolsep}{5pt}
\begin{tabular}{lcccccc}
\toprule
 & \multicolumn{3}{c}{Newly Requested Workers per 1,000} & \multicolumn{3}{c}{New Filing Firms per 1,000} \\
\cmidrule(lr){2-4} \cmidrule(lr){5-7}
 & (1) & (2) & (3) & (4) & (5) & (6) \\
 & High Tech & Low Tech & Full Sample & High Tech & Low Tech & Full Sample \\
\midrule
Mort. Pay. Wedge $\times$ Post & 0.008 & 0.027*** & 0.009* & 0.0019*** & 0.0025*** & 0.0017*** \\
 & (0.005) & (0.008) & (0.005) & (0.0006) & (0.0007) & (0.0005) \\[0.5em]
Mort. Pay. Wedge $\times$ Post $\times$ Tech Sh. &  &  & $-$0.022*** &  &  & $-$0.0011* \\
 &  &  & (0.008) &  &  & (0.0006) \\[0.5em]
Post $\times$ Tech Sh. &  &  & 4.468** &  &  & 0.222* \\
 &  &  & (1.725) &  &  & (0.123) \\[0.5em]
\midrule
Observations & 2,896 & 2,579 & 5,475 & 2,896 & 2,579 & 5,475 \\
CZ and Year Fixed Effects & Yes & Yes & Yes & Yes & Yes & Yes \\
Time-Varying Controls & Yes & Yes & Yes & Yes & Yes & Yes \\
\bottomrule
\end{tabular}
\begin{minipage}{\linewidth}
\vspace{0.5em}
\footnotesize This table replicates Table~\ref{tab:h1b_cz} by local tech employment intensity. Columns (1) and (4) restrict to commuting zones at or above the median 2017 to 2019 CZ tech employment share; columns (2) and (5) to those below; columns (3) and (6) use the full sample with the Mort. Pay. Wedge $\times$ Post $\times$ Tech Share triple interaction and the Post $\times$ Tech Share interaction added. Tech Share is the 2017 to 2019 CZ tech employment share in percentage points, demeaned by the full-sample mean. The rest of the specification follows from Tables~\ref{tab:migration_costin} and~\ref{tab:h1b_cz}. *** p$<$0.01, ** p$<$0.05, * p$<$0.10.
\end{minipage}
\end{table}
\end{landscape}

%% file: new_tables/tab_h1b_cz_soc_decomposition.tex
\begin{table}[htbp]
\centering
\caption{\textbf{Decomposition of the H-1B Response by SOC Demand Shock}}
\label{tab:h1b_cz_soc_bartik_decomposition}
\begin{tabular}{lcccc}
\toprule
& (1)       & (2)       & (3)  & (4)     \\
& Newly     & Total     & New  & New     \\
& Requested & Requested & LCAs & Filing  \\
& Workers   & Workers   &      & Firms   \\
\midrule
2024 Mort. Pay. $\times$ Post $\times$ Bartik     & 0.068**     & 0.176***    & 0.014       & 0.0129***  \\
                                                        & (0.033)     & (0.067)     & (0.013)     & (0.0042)   \\
[0.5em]
Locked-In Origin Pay. $\times$ Post $\times$ Bartik  & $-$0.045*** & $-$0.094*** & $-$0.022*** & 0.0006     \\
                                                        & (0.013)     & (0.020)     & (0.006)     & (0.0020)   \\
\midrule
Observations               & 43,889 & 43,889 & 43,889 & 43,889 \\
\cmidrule{1-5}
CZ $\times$ Year FE        & Yes & Yes & Yes & Yes \\
CZ $\times$ SOC FE         & Yes & Yes & Yes & Yes \\
SOC $\times$ Year FE       & Yes & Yes & Yes & Yes \\
\bottomrule
\end{tabular}
\vspace{0.5em}
\begin{minipage}{\linewidth}
\footnotesize This table replicates Table~\ref{tab:h1b_cz_soc_bartik} with $P^{new}_{d}$ and $WOP_{d}$ entered jointly in place of $MPW^{in}_{d}$, defined as in Table~\ref{tab:migration_decomposition}. The reported rows are the triple interactions of each component with Post and the Bartik shock. The rest of the specification follows from Table~\ref{tab:h1b_cz_soc_bartik}. *** p$<$0.01, ** p$<$0.05, * p$<$0.10.
\end{minipage}
\end{table}

%% file: new_tables/tab_bartik_bins.tex
\begin{table}[htbp]
\centering
\caption{\textbf{Mortgage Payment Wedge Effects Across Positive Bartik Demand Bins}}
\label{tab:positive_bartik_bins}
\begin{tabular}{lcccc}
\toprule
& (1) & (2) & (3) & (4) \\
& Newly     & Total     &      &            \\
& Requested & Requested & New  & New Filing \\
& Workers   & Workers   & LCAs & Firms      \\
\midrule
Mort. Pay. Wedge $\times$ Post $\times \cdots$ & & & & \\[0.4em]
\hspace{1.2em} Bartik$^{+}$ Low  & $-$0.024 & $-$0.199* & $-$0.023 & $-$0.0228* \\
                        & (0.037)  & (0.106)   & (0.024)  & (0.0119)   \\
[0.5em]
\hspace{1.2em} Bartik$^{+}$ Mid  & 0.054    & 0.047     & 0.039*** & 0.0111   \\
                        & (0.058)  & (0.086)   & (0.015)  & (0.0077) \\
[0.5em]
\hspace{1.2em} Bartik$^{+}$ High & 0.063**  & 0.135**   & 0.026*   & 0.0013   \\
                        & (0.030)  & (0.054)   & (0.014)  & (0.0044) \\
[0.5em]
\midrule
Joint Test $p$-value & 0.122  & 0.003  & 0.040  & 0.069  \\
Observations         & 43,889 & 43,889 & 43,889 & 43,889 \\
CZ $\times$ Year FE  & Yes & Yes & Yes & Yes \\
CZ $\times$ SOC FE   & Yes & Yes & Yes & Yes \\
SOC $\times$ Year FE & Yes & Yes & Yes & Yes \\
\bottomrule
\end{tabular}
\vspace{0.5em}
\begin{minipage}{\linewidth}
\footnotesize This table replicates Table~\ref{tab:h1b_cz_soc_bartik} with the Bartik shock entered as bins rather than a continuous measure. Observations with a nonpositive Bartik shock form the reference group; those with a positive shock are split into terciles (low, mid, high). Each coefficient is the triple interaction of the wedge and Post with the indicated positive-Bartik tercile, relative to the nonpositive reference. The joint test $p$-value is for the null that the three triple interactions are jointly zero. The rest of the specification follows from Table~\ref{tab:h1b_cz_soc_bartik}. *** p$<$0.01, ** p$<$0.05, * p$<$0.10.
\end{minipage}
\end{table}

%% file: new_tables/tab_uscis_decomposition.tex
\begin{table}[htbp]
\centering
\caption{\textbf{Decomposition of the USCIS H-1B Response to the In-Migration Mortgage-Payment Wedge}}
\label{tab:uscis_decomposition}
\begin{tabular}{lcccc}
\toprule
& (1)      & (2)      & (3)       & (4)    \\
& Newly    & Total    & New       & Filing \\
& Approved & Approved & Petitions & Firms  \\
& Workers  & Workers  &           &        \\
\midrule
2024 Mortgage Payment $\times$ Post     & 0.0071      & 0.0070      & 0.0076      & 0.0030*       \\
                                        & (0.0061)    & (0.0168)    & (0.0056)    & (0.0018)      \\
[0.5em]
Locked-In Origin Payment $\times$ Post  & $-$0.0039   & $-$0.0079*  & $-$0.0045   & $-$0.0027***  \\
                                        & (0.0030)    & (0.0045)    & (0.0030)    & (0.0005)      \\
\midrule
Observations               & 4,888 & 4,888 & 4,888 & 4,888 \\
\cmidrule{1-5}
CZ and Year Fixed Effects  & Yes & Yes & Yes & Yes \\
Time-Varying Controls      & Yes & Yes & Yes & Yes \\
\bottomrule
\end{tabular}
\vspace{0.5em}
\begin{minipage}{\linewidth}
\footnotesize This table replicates Table~\ref{tab:uscis} with $P^{new}_{d}$ and $WOP_{d}$ entered jointly in place of $MPW^{in}_{d}$, defined as in Table~\ref{tab:migration_decomposition}. The rest of the specification follows from Table~\ref{tab:uscis}. *** p$<$0.01, ** p$<$0.05, * p$<$0.10.
\end{minipage}
\end{table}

%% file: new_tables/tab_iv_mpw.tex
\begin{table}[htbp]
\centering
\caption{\textbf{Instrumenting the Mortgage Payment Wedge}}
\label{tab:iv_mpw}
\begin{tabular}{lccc}
\toprule
\multicolumn{4}{c}{\textit{Panel A: Migration Outcomes (per 1,000 population)}} \\
\cmidrule(lr){2-4}
& (1) & (2) & (3) \\
& College      & College & College \\
& In-Migration & Owners  & Renters \\
\midrule
Mortgage Payment Wedge $\times$ Post & $-$0.008 & $-$0.061* & 0.053   \\
                                     & (0.050)  & (0.033)   & (0.033) \\
\midrule
CZ and Year Fixed Effects   & Yes   & Yes   & Yes   \\
Time-Varying Controls       & Yes   & Yes   & Yes   \\
Observations                & 5,472 & 5,472 & 5,472 \\
\midrule
\multicolumn{4}{c}{\textit{Panel B: H-1B Outcomes (per 1,000 employed)}} \\
\cmidrule(lr){2-4}
& (1) & (2) & (3) \\
& Newly Requested & New  & New Filing \\
& Workers         & LCAs & Firms      \\
\midrule
Mortgage Payment Wedge $\times$ Post & 0.046** & 0.008   & 0.0051*** \\
                                     & (0.019) & (0.009) & (0.0020)  \\
\midrule
CZ and Year Fixed Effects   & Yes   & Yes   & Yes   \\
Time-Varying Controls       & Yes   & Yes   & Yes   \\
Observations                & 5,472 & 5,472 & 5,472 \\
\bottomrule
\end{tabular}
\vspace{0.5em}
\begin{minipage}{\linewidth}
\footnotesize Two-stage least squares estimates on the commuting-zone-year panel. The Mortgage Payment Wedge $\times$ Post is the endogenous regressor, instrumented by a predicted wedge equal to the 2024 Mortgage Payment minus the predicted Locked-In Origin Payment of Table~\ref{tab:iv_wop}. The first-stage coefficient is 0.68 and the cluster-robust Kleibergen-Paap first-stage $F$ is 70.06, on 5,472 commuting-zone-year cells across 684 commuting zones. The rest of the specification follows from Tables~\ref{tab:migration_costin} and~\ref{tab:h1b_cz}. *** p$<$0.01, ** p$<$0.05, * p$<$0.10.
\end{minipage}
\end{table}

%% file: new_tables/tab_wop_topk.tex
\begin{table}[htbp]
\centering
\caption{\textbf{Robustness to Truncating Feeder Origins (Top-$K$)}}
\label{tab:robust_wop_topk}
\begin{tabular}{lcc}
\toprule
& (1) & (2) \\
& College      & Newly Requested \\
& In-Migration & H-1B Workers \\
\midrule
Baseline (all feeder origins) & $-$0.059*** & 0.018*** \\
                              & (0.015)     & (0.006)  \\
[0.5em]
Top-1                         & $-$0.038*** & 0.015*** \\
                              & (0.011)     & (0.005)  \\
[0.5em]
Top-3                         & $-$0.052*** & 0.017*** \\
                              & (0.013)     & (0.005)  \\
[0.5em]
Top-5                         & $-$0.057*** & 0.017*** \\
                              & (0.014)     & (0.005)  \\
[0.5em]
Top-10                        & $-$0.057*** & 0.017*** \\
                              & (0.014)     & (0.005)  \\
[0.5em]
Top-20                        & $-$0.059*** & 0.018*** \\
                              & (0.015)     & (0.005)  \\
\midrule
Observations                  & 5,475       & 5,475    \\
CZ and Year Fixed Effects     & Yes         & Yes      \\
Time-Varying Controls         & Yes         & Yes      \\
\bottomrule
\end{tabular}
\vspace{0.5em}
\begin{minipage}{\linewidth}
\footnotesize This table replicates Tables~\ref{tab:migration_costin} and~\ref{tab:h1b_cz} with the wedge rebuilt from the top $K$ feeder origins by pre-shock IRS inflows, $WOP^K_d = \sum_{o \in \text{Top-}K(d)} \omega^K_{od} P^{old}_o$ (weights renormalized to one), and the truncated wedge $P^{new}_d - WOP^K_d$ interacted with Post. Median coverage of inflows by the truncated set is 41 percent at $K=1$, 75 percent at $K=3$, 92 percent at $K=5$, and 100 percent at $K \geq 10$. The rest of the specification follows from those tables. *** p$<$0.01, ** p$<$0.05, * p$<$0.10.
\end{minipage}
\end{table}

%% file: new_tables/tab_drop_top.tex
\begin{table}[htbp]
\centering
\caption{\textbf{Robustness to Dropping Top H-1B Commuting Zones}}
\label{tab:drop_top_robustness}
\begin{tabular}{lcccc}
\toprule
\multicolumn{5}{l}{\textit{Panel A: MPW$^{in}_c \times$ Post Estimates, CZ-Year Migration}} \\
\midrule
& (1) & (2) & & \\
& College      & College &  &  \\
& In-Migration & Owners  &  &  \\
\midrule
Drop Top 25  & $-$0.061*** & $-$0.060*** &  &  \\
             & (0.015)     & (0.010)     &  &  \\
[0.5em]
Drop Top 50  & $-$0.061*** & $-$0.059*** &  &  \\
             & (0.015)     & (0.010)     &  &  \\
[0.5em]
Drop Top 100 & $-$0.066*** & $-$0.063*** &  &  \\
             & (0.015)     & (0.010)     &  &  \\
[0.5em]
\midrule
\multicolumn{5}{l}{\textit{Panel B: MPW$^{in}_c \times$ Post Estimates, CZ-Year H-1B}} \\
\midrule
& (1) & (2) & (3) & (4) \\
& Newly Requested & Total Requested & New  & New Filing \\
& Workers         & Workers         & LCAs & Firms      \\
\midrule
Drop Top 25  & 0.021*** & 0.032*** & 0.011*** & 0.0021*** \\
             & (0.005)  & (0.009)  & (0.004)  & (0.0005)  \\
[0.5em]
Drop Top 50  & 0.021*** & 0.034*** & 0.011*** & 0.0022*** \\
             & (0.005)  & (0.009)  & (0.004)  & (0.0005)  \\
[0.5em]
Drop Top 100 & 0.021*** & 0.032*** & 0.012*** & 0.0022*** \\
             & (0.006)  & (0.009)  & (0.004)  & (0.0005)  \\
\bottomrule
\end{tabular}
\vspace{0.5em}
\begin{minipage}{\linewidth}
\footnotesize Panels A and B replicate Tables~\ref{tab:migration_costin} and~\ref{tab:h1b_cz} after dropping the 25, 50, or 100 commuting zones with the highest H-1B activity (rows). The rest of the specification follows from those tables. *** p$<$0.01, ** p$<$0.05, * p$<$0.10.
\end{minipage}
\end{table}

%% file: new_tables/tab_drop_covid.tex
\begin{landscape}
\begin{table}[htbp]
\centering
\caption{\textbf{Robustness to Excluding the Pandemic Years}}
\label{tab:drop_covid}
\begin{tabular}{lccccc}
\toprule
\multicolumn{6}{l}{\textit{Panel A: Migration Outcomes (per 1,000 population)}} \\
\cmidrule(lr){2-6}
& (1) & (2) & (3) & (4) & (5) \\
& College      & College & College & Non-College  & College     \\
& In-Migration & Owners  & Renters & In-Migration & From Abroad \\
\midrule
MPW$^{in}_c$ $\times$ Post & $-$0.074*** & $-$0.065*** & $-$0.009 & $-$0.108*  & 0.016*** \\
                           & (0.018)     & (0.012)     & (0.012)  & (0.062)    & (0.004)  \\
\midrule
CZ and Year Fixed Effects  & Yes   & Yes   & Yes   & Yes   & Yes   \\
Time-Varying Controls      & Yes   & Yes   & Yes   & Yes   & Yes   \\
Observations               & 4,107 & 4,107 & 4,107 & 4,107 & 4,107 \\
\midrule
\multicolumn{6}{l}{\textit{Panel B: H-1B Outcomes (per 1,000 employed)}} \\
\cmidrule(lr){2-5}
& (1) & (2) & (3) & (4) & \\
& Newly Requested & Total Requested & New  & New Filing &  \\
& Workers         & Workers         & LCAs & Firms      &  \\
\midrule
MPW$^{in}_c$ $\times$ Post & 0.022*** & 0.035*** & 0.013*** & 0.0028*** &  \\
                           & (0.006)  & (0.010)  & (0.004)  & (0.0006)  &  \\
\midrule
CZ and Year Fixed Effects  & Yes   & Yes   & Yes   & Yes   & \\
Time-Varying Controls      & Yes   & Yes   & Yes   & Yes   & \\
Observations               & 4,107 & 4,107 & 4,107 & 4,107 & \\
\bottomrule
\end{tabular}
\vspace{0.5em}
\begin{minipage}{\linewidth}
\footnotesize This table replicates Tables~\ref{tab:migration_costin} and~\ref{tab:h1b_cz} excluding the pandemic years 2020 and 2021, so the sample covers 2017 to 2019 and 2022 to 2024. The rest of the specification follows from those tables. *** p$<$0.01, ** p$<$0.05, * p$<$0.10.
\end{minipage}
\end{table}
\end{landscape}

%% file: new_tables/tab_drop_2024.tex
\begin{landscape}
\begin{table}[htbp]
\centering
\caption{\textbf{Robustness to Excluding 2024}}
\label{tab:drop_2024}
\begin{tabular}{lccccc}
\toprule
\multicolumn{6}{c}{\textit{Panel A: Migration Outcomes (per 1,000 population)}} \\
\cmidrule(lr){2-6}
& (1) & (2) & (3) & (4) & (5) \\
& College      & College & College & Non-College  & College     \\
& In-Migration & Owners  & Renters & In-Migration & From Abroad \\
\midrule
MPW$^{in}_c$ $\times$ Post & $-$0.041** & $-$0.054*** & 0.013   & $-$0.046 & 0.016*** \\
                           & (0.016)    & (0.011)     & (0.010) & (0.053)  & (0.004)  \\
\midrule
CZ and Year Fixed Effects  & Yes   & Yes   & Yes   & Yes   & Yes   \\
Time-Varying Controls      & Yes   & Yes   & Yes   & Yes   & Yes   \\
Observations               & 4,791 & 4,791 & 4,791 & 4,791 & 4,791 \\
\midrule
\multicolumn{6}{c}{\textit{Panel B: H-1B Outcomes (per 1,000 employed)}} \\
\cmidrule(lr){2-5}
& (1) & (2) & (3) & (4) & \\
& Newly Requested & Total Requested & New  & New Filing &  \\
& Workers         & Workers         & LCAs & Firms      &  \\
\midrule
MPW$^{in}_c$ $\times$ Post & 0.012** & 0.010   & 0.007** & 0.0015*** &  \\
                           & (0.006) & (0.008) & (0.003) & (0.0004)  &  \\
\midrule
CZ and Year Fixed Effects  & Yes   & Yes   & Yes   & Yes   & \\
Time-Varying Controls      & Yes   & Yes   & Yes   & Yes   & \\
Observations               & 4,791 & 4,791 & 4,791 & 4,791 & \\
\bottomrule
\end{tabular}
\vspace{0.5em}
\begin{minipage}{\linewidth}
\footnotesize This table replicates Tables~\ref{tab:migration_costin} and~\ref{tab:h1b_cz} excluding 2024, so the sample covers 2017 to 2023 and Post indicates 2022 and 2023. The rest of the specification follows from those tables. *** p$<$0.01, ** p$<$0.05, * p$<$0.10.
\end{minipage}
\end{table}
\end{landscape}

%% file: new_tables/tab_hpi.tex
\begin{landscape}
\begin{table}[htbp]
\centering
\caption{\textbf{Robustness to Including House Price Controls}}
\label{tab:hpi_controls}
\begin{tabular}{lccccc}
\toprule
\multicolumn{6}{c}{\textit{Panel A: Migration Outcomes (per 1,000 population)}} \\
\cmidrule(lr){2-6}
& (1) & (2) & (3) & (4) & (5) \\
& College      & College & College & Non-College  & College     \\
& In-Migration & Owners  & Renters & In-Migration & From Abroad \\
\midrule
MPW$^{in}_c$ $\times$ Post & $-$0.054*** & $-$0.058*** & 0.004   & $-$0.096*  & 0.011*** \\
                           & (0.015)     & (0.010)     & (0.010) & (0.049)    & (0.004)  \\
\midrule
CZ and Year Fixed Effects  & Yes   & Yes   & Yes   & Yes   & Yes   \\
Time-Varying Controls      & Yes   & Yes   & Yes   & Yes   & Yes   \\
Observations               & 5,302 & 5,302 & 5,302 & 5,302 & 5,302 \\
\midrule
\multicolumn{6}{c}{\textit{Panel B: H-1B Outcomes (per 1,000 employed)}} \\
\cmidrule(lr){2-5}
& (1) & (2) & (3) & (4) & \\
& Newly Requested & Total Requested & New  & New Filing &  \\
& Workers         & Workers         & LCAs & Firms      &  \\
\midrule
MPW$^{in}_c$ $\times$ Post & 0.0070** & 0.0086   & 0.0019*  & 0.0012*** &  \\
                           & (0.0029) & (0.0067) & (0.0011) & (0.0004)  &  \\
\midrule
CZ and Year Fixed Effects  & Yes   & Yes   & Yes   & Yes   & \\
Time-Varying Controls      & Yes   & Yes   & Yes   & Yes   & \\
Observations               & 5,302 & 5,302 & 5,302 & 5,302 & \\
\bottomrule
\end{tabular}
\vspace{0.5em}
\begin{minipage}{\linewidth}
\footnotesize This table replicates Tables~\ref{tab:migration_costin} and~\ref{tab:h1b_cz} with two FHFA House Price Index controls added (the log price index and its annual percent change). The sample is restricted to the 666 commuting zones with non-missing house price data. The rest of the specification follows from those tables. *** p$<$0.01, ** p$<$0.05, * p$<$0.10.
\end{minipage}
\end{table}
\end{landscape}

%% file: new_tables/tab_alt_functions.tex
\begin{table}[htbp]
\centering
\caption{\textbf{Robustness to Alternative Functional Forms}}
\label{tab:alt_functional_forms}
\begin{tabular}{lcccc}
\toprule
& (1) & (2) & (3) & (4) \\
& Newly Requested & Total Requested & New  & New Filing \\
& Workers         & Workers         & LCAs & Firms      \\
\midrule
\multicolumn{5}{c}{\textit{Panel A: $\log(0.1 + y)$}} \\
\midrule
MPW$^{in}_c$ $\times$ Post & 0.027*** & 0.023*** & 0.022*** & 0.017*** \\
                           & (0.006)  & (0.005)  & (0.005)  & (0.004)  \\
Observations               & 5,475    & 5,475    & 5,475    & 5,475    \\
\midrule
\multicolumn{5}{c}{\textit{Panel B: $\log(1 + y)$}} \\
\midrule
MPW$^{in}_c$ $\times$ Post & 0.017*** & 0.017*** & 0.012*** & 0.008*** \\
                           & (0.004)  & (0.004)  & (0.003)  & (0.002)  \\
Observations               & 5,475    & 5,475    & 5,475    & 5,475    \\
\midrule
\multicolumn{5}{c}{\textit{Panel C: Inverse Hyperbolic Sine}} \\
\midrule
MPW$^{in}_c$ $\times$ Post & 0.020*** & 0.019*** & 0.015*** & 0.009*** \\
                           & (0.004)  & (0.004)  & (0.004)  & (0.002)  \\
Observations               & 5,475    & 5,475    & 5,475    & 5,475    \\
\midrule
\multicolumn{5}{c}{\textit{Panel D: Negative Binomial Fixed Effects}} \\
\midrule
MPW$^{in}_c$ $\times$ Post & 0.0084*** & 0.0068*** & $-$0.0023 & 0.0018   \\
                           & (0.0026)  & (0.0022)  & (0.0021)  & (0.0018) \\
Observations               & 5,384     & 5,451     & 5,384     & 5,384    \\
\midrule
\multicolumn{5}{c}{\textit{Panel E: Long Difference, 2024 $-$ 2019}} \\
\midrule
MPW$^{in}_c$               & 0.040*** & 0.077*** & 0.017*** & 0.003*** \\
                           & (0.013)  & (0.020)  & (0.006)  & (0.001)  \\
Observations               & 684      & 684      & 684      & 684      \\
\bottomrule
\end{tabular}
\vspace{0.5em}
\begin{minipage}{\linewidth}
\footnotesize This table replicates Table~\ref{tab:h1b_cz} under alternative functional forms. Panels A to C apply $\log(0.1+y)$, $\log(1+y)$, and the inverse hyperbolic sine to the H-1B count outcome. Panel D reports conditional fixed-effects negative binomial estimates with log commuting-zone employment as an offset and year indicators, dropping commuting zones with all-zero outcomes so the sample varies across columns. Panel E regresses the 2019-to-2024 long difference per 1,000 employed workers on $MPW^{in}_c$ and 2019 commuting-zone controls in a cross-section. Standard errors are clustered at the commuting-zone level in Panels A to D and heteroskedasticity-robust in Panel E. The rest of the specification follows from Table~\ref{tab:h1b_cz}. *** p$<$0.01, ** p$<$0.05, * p$<$0.10.
\end{minipage}
\end{table}

%% file: new_tables/tab_ipeds.tex
\begin{landscape}
\begin{table}[htbp]
\centering
\caption{\textbf{Placebo Test Using IPEDS College Enrollment}}
\label{tab:ipeds_placebo}
\begin{tabular}{lccccc}
\toprule
& \multicolumn{4}{c}{Per 1,000 Population} & \\
\cmidrule(lr){2-5}
& (1) & (2) & (3) & (4) & (5) \\
& Foreign    & Foreign       & Foreign  & Total      & Foreign Share \\
& Enrollment & Undergraduate & Graduate & Enrollment & of Enrollment \\
\midrule
MPW$^{in}_c$ $\times$ Post & $-$0.013 & $-$0.008 & $-$0.005 & $-$0.048 & 0.0000   \\
                           & (0.017)  & (0.011)  & (0.014)  & (0.044)  & (0.0001) \\
\midrule
CZ and Year Fixed Effects  & Yes   & Yes   & Yes   & Yes   & Yes   \\
Time-Varying Controls      & Yes   & Yes   & Yes   & Yes   & Yes   \\
Observations               & 5,475 & 5,475 & 5,475 & 5,475 & 5,475 \\
\bottomrule
\end{tabular}
\vspace{0.5em}
\begin{minipage}{\linewidth}
\footnotesize This table replicates Table~\ref{tab:migration_costin} with IPEDS college enrollment outcomes (foreign total, foreign undergraduate, foreign graduate, total enrollment, and foreign share) in place of the migration outcomes, as a placebo test. The rest of the specification follows from Table~\ref{tab:migration_costin}. *** p$<$0.01, ** p$<$0.05, * p$<$0.10.
\end{minipage}
\end{table}
\end{landscape}

%% file: references.bib
@article{mayda_effect_2018,
	title = {The effect of the {H}-{1B} quota on the employment and selection of foreign-born labor},
	volume = {108},
	issn = {00142921},
	url = {https://linkinghub.elsevier.com/retrieve/pii/S0014292118300965},
	doi = {10.1016/j.euroecorev.2018.06.010},
	language = {en},
	urldate = {2026-05-26},
	journal = {European Economic Review},
	author = {Mayda, Anna Maria and Ortega, Francesc and Peri, Giovanni and Shih, Kevin and Sparber, Chad},
	month = sep,
	year = {2018},
	pages = {105--128},
}

@techreport{olney_determinants_2024,
	address = {Cambridge, MA},
	title = {The {Determinants} of {Declining} {Internal} {Migration}},
	url = {http://www.nber.org/papers/w32123.pdf},
	language = {en},
	number = {w32123},
	urldate = {2026-05-26},
	institution = {National Bureau of Economic Research},
	author = {Olney, William and Thompson, Owen},
	month = feb,
	year = {2024},
	doi = {10.3386/w32123},
	pages = {w32123},
}

@article{ferreira_housing_2010,
	title = {Housing busts and household mobility},
	volume = {68},
	issn = {00941190},
	url = {https://linkinghub.elsevier.com/retrieve/pii/S0094119009000886},
	doi = {10.1016/j.jue.2009.10.007},
	language = {en},
	number = {1},
	urldate = {2026-05-26},
	journal = {Journal of Urban Economics},
	author = {Ferreira, Fernando and Gyourko, Joseph and Tracy, Joseph},
	month = jul,
	year = {2010},
	pages = {34--45},
}

@article{blanchard_regional_1992,
	title = {Regional {Evolutions}},
	volume = {1992},
	issn = {00072303},
	url = {https://www.jstor.org/stable/2534556?origin=crossref},
	doi = {10.2307/2534556},
	number = {1},
	urldate = {2026-05-26},
	journal = {Brookings Papers on Economic Activity},
	author = {Blanchard, Olivier Jean and Katz, Lawrence F. and Hall, Robert E. and Eichengreen, Barry},
	year = {1992},
	pages = {1},
}

@article{notowidigdo_incidence_2020,
	title = {The {Incidence} of {Local} {Labor} {Demand} {Shocks}},
	volume = {38},
	issn = {0734-306X, 1537-5307},
	url = {https://www.journals.uchicago.edu/doi/10.1086/706048},
	doi = {10.1086/706048},
	language = {en},
	number = {3},
	urldate = {2026-05-26},
	journal = {Journal of Labor Economics},
	author = {Notowidigdo, Matthew J.},
	month = jul,
	year = {2020},
	pages = {687--725},
}

@article{peri_stem_2015,
	title = {{STEM} {Workers}, {H}-{1B} {Visas}, and {Productivity} in {US} {Cities}},
	volume = {33},
	issn = {0734-306X, 1537-5307},
	url = {https://www.journals.uchicago.edu/doi/10.1086/679061},
	doi = {10.1086/679061},
	language = {en},
	number = {S1},
	urldate = {2026-05-26},
	journal = {Journal of Labor Economics},
	author = {Peri, Giovanni and Shih, Kevin and Sparber, Chad},
	month = jul,
	year = {2015},
	pages = {S225--S255},
}

@article{cadena_immigrants_2016,
	title = {Immigrants {Equilibrate} {Local} {Labor} {Markets}: {Evidence} from the {Great} {Recession}},
	volume = {8},
	issn = {1945-7782, 1945-7790},
	shorttitle = {Immigrants {Equilibrate} {Local} {Labor} {Markets}},
	url = {https://pubs.aeaweb.org/doi/10.1257/app.20140095},
	doi = {10.1257/app.20140095},
	language = {en},
	number = {1},
	urldate = {2026-05-26},
	journal = {American Economic Journal: Applied Economics},
	author = {Cadena, Brian C. and Kovak, Brian K.},
	month = jan,
	year = {2016},
	pages = {257--290},
}

@article{beraja_regional_2019,
	title = {Regional {Heterogeneity} and the {Refinancing} {Channel} of {Monetary} {Policy}*},
	volume = {134},
	copyright = {https://academic.oup.com/journals/pages/open\_access/funder\_policies/chorus/standard\_publication\_model},
	issn = {0033-5533, 1531-4650},
	url = {https://academic.oup.com/qje/article/134/1/109/5089981},
	doi = {10.1093/qje/qjy021},
	language = {en},
	number = {1},
	urldate = {2026-05-20},
	journal = {The Quarterly Journal of Economics},
	author = {Beraja, Martin and Fuster, Andreas and Hurst, Erik and Vavra, Joseph},
	month = feb,
	year = {2019},
	pages = {109--183},
}

@article{carlino_differential_1998,
	title = {The {Differential} {Regional} {Effects} of {Monetary} {Policy}},
	volume = {80},
	issn = {0034-6535, 1530-9142},
	url = {https://direct.mit.edu/rest/article/80/4/572-587/57112},
	doi = {10.1162/003465398557843},
	language = {en},
	number = {4},
	urldate = {2026-05-20},
	journal = {Review of Economics and Statistics},
	author = {Carlino, Gerald and DeFina, Robert},
	month = nov,
	year = {1998},
	pages = {572--587},
}

@article{romer_new_2004,
	title = {A {New} {Measure} of {Monetary} {Shocks}: {Derivation} and {Implications}},
	volume = {94},
	issn = {0002-8282},
	shorttitle = {A {New} {Measure} of {Monetary} {Shocks}},
	url = {https://pubs.aeaweb.org/doi/10.1257/0002828042002651},
	doi = {10.1257/0002828042002651},
	language = {en},
	number = {4},
	urldate = {2026-05-20},
	journal = {American Economic Review},
	author = {Romer, Christina D and Romer, David H},
	month = sep,
	year = {2004},
	pages = {1055--1084},
}

@article{nakamura_identification_2018,
	title = {Identification in {Macroeconomics}},
	volume = {32},
	issn = {0895-3309},
	url = {https://pubs.aeaweb.org/doi/10.1257/jep.32.3.59},
	doi = {10.1257/jep.32.3.59},
	language = {en},
	number = {3},
	urldate = {2026-05-20},
	journal = {Journal of Economic Perspectives},
	author = {Nakamura, Emi and Steinsson, Jón},
	month = aug,
	year = {2018},
	pages = {59--86},
}

@article{dingel_how_2020,
	title = {How many jobs can be done at home?},
	volume = {189},
	issn = {00472727},
	url = {https://linkinghub.elsevier.com/retrieve/pii/S0047272720300992},
	doi = {10.1016/j.jpubeco.2020.104235},
	language = {en},
	urldate = {2026-05-14},
	journal = {Journal of Public Economics},
	author = {Dingel, Jonathan I. and Neiman, Brent},
	month = sep,
	year = {2020},
	pages = {104235},
}

@article{kerr_supply_2010,
	title = {The {Supply} {Side} of {Innovation}: {H}‐{1B} {Visa} {Reforms} and {U}.{S}. {Ethnic} {Invention}},
	volume = {28},
	issn = {0734-306X, 1537-5307},
	shorttitle = {The {Supply} {Side} of {Innovation}},
	url = {https://www.journals.uchicago.edu/doi/10.1086/651934},
	doi = {10.1086/651934},
	language = {en},
	number = {3},
	urldate = {2026-05-14},
	journal = {Journal of Labor Economics},
	author = {Kerr, William R. and Lincoln, William F.},
	month = jul,
	year = {2010},
	pages = {473--508},
}

@article{kerr_skilled_2015,
	title = {Skilled {Immigration} and the {Employment} {Structures} of {US} {Firms}},
	volume = {33},
	issn = {0734-306X, 1537-5307},
	url = {https://www.journals.uchicago.edu/doi/10.1086/678986},
	doi = {10.1086/678986},
	language = {en},
	number = {S1},
	urldate = {2026-05-14},
	journal = {Journal of Labor Economics},
	author = {Kerr, Sari Pekkala and Kerr, William R. and Lincoln, William F.},
	month = jul,
	year = {2015},
	pages = {S147--S186},
}

@article{doran_effects_2022,
	title = {The {Effects} of {High}-{Skilled} {Immigration} {Policy} on {Firms}: {Evidence} from {Visa} {Lotteries}},
	volume = {130},
	issn = {0022-3808, 1537-534X},
	shorttitle = {The {Effects} of {High}-{Skilled} {Immigration} {Policy} on {Firms}},
	url = {https://www.journals.uchicago.edu/doi/10.1086/720467},
	doi = {10.1086/720467},
	language = {en},
	number = {10},
	urldate = {2026-05-14},
	journal = {Journal of Political Economy},
	author = {Doran, Kirk and Gelber, Alexander and Isen, Adam},
	month = oct,
	year = {2022},
	pages = {2501--2533},
}

@article{tolbert_us_1996,
	title = {{US} commuting zones and labor market areas: {A} 1990 update},
	author = {Tolbert, Charles M and Sizer, Molly},
	year = {1996},
}

@article{becker_persecution_2024,
	title = {Persecution and {Escape}: {Professional} {Networks} and {High}-{Skilled} {Emigration} from {Nazi} {Germany}},
	volume = {16},
	issn = {1945-7782, 1945-7790},
	shorttitle = {Persecution and {Escape}},
	url = {https://pubs.aeaweb.org/doi/10.1257/app.20220278},
	doi = {10.1257/app.20220278},
	language = {en},
	number = {3},
	urldate = {2026-01-08},
	journal = {American Economic Journal: Applied Economics},
	author = {Becker, Sascha O. and Lindenthal, Volker and Mukand, Sharun W. and Waldinger, Fabian},
	month = jul,
	year = {2024},
	pages = {1--43},
}

@article{fasani_economics_2020,
	title = {The economics of migration: {Labour} market impacts and migration policies},
	volume = {67},
	issn = {09275371},
	shorttitle = {The economics of migration},
	url = {https://linkinghub.elsevier.com/retrieve/pii/S0927537120301330},
	doi = {10.1016/j.labeco.2020.101929},
	language = {en},
	urldate = {2026-01-08},
	journal = {Labour Economics},
	author = {Fasani, Francesco and Llull, Joan and Tealdi, Cristina},
	month = dec,
	year = {2020},
	pages = {101929},
}

@article{parey_selection_2017,
	title = {The selection of high-skilled emigrants},
	volume = {99},
	issn = {0034-6535},
	number = {5},
	journal = {Review of Economics and Statistics},
	publisher = {MIT Press One Rogers Street, Cambridge, MA 02142-1209, USA journals-info …},
	author = {Parey, Matthias and Ruhose, Jens and Waldinger, Fabian and Netz, Nicolai},
	year = {2017},
	pages = {776--792},
}

@article{bound_recruitment_2015,
	title = {Recruitment of foreigners in the market for computer scientists in the {United} {States}},
	volume = {33},
	issn = {0734-306X},
	number = {S1},
	journal = {Journal of labor economics},
	publisher = {University of Chicago Press Chicago, IL},
	author = {Bound, John and Braga, Breno and Golden, Joseph M and Khanna, Gaurav},
	year = {2015},
	pages = {S187--S223},
}

@article{mahajan_immigration_2024,
	title = {Immigration and business dynamics: {Evidence} from us firms},
	volume = {22},
	issn = {1542-4766},
	number = {6},
	journal = {Journal of the European Economic Association},
	publisher = {Oxford University Press},
	author = {Mahajan, Parag},
	year = {2024},
	pages = {2827--2869},
}

@incollection{hanson_high-skilled_2017,
	title = {High-skilled immigration and the rise of {STEM} occupations in {US} employment},
	booktitle = {Education, skills, and technical change: {Implications} for future {US} {GDP} {Growth}},
	publisher = {University of Chicago Press},
	author = {Hanson, Gordon H and Slaughter, Matthew J},
	year = {2017},
	pages = {465--494},
}

@techreport{clemens_effect_2022,
	title = {The effect of low-skill immigration restrictions on {US} firms and workers: {Evidence} from a randomized lottery},
	institution = {National Bureau of Economic Research},
	author = {Clemens, Michael A and Lewis, Ethan G},
	year = {2022},
}

@article{glennon_how_2024,
	title = {How do restrictions on high-skilled immigration affect offshoring? {Evidence} from the {H}-{1B} program},
	volume = {70},
	issn = {0025-1909},
	number = {2},
	journal = {Management Science},
	publisher = {INFORMS},
	author = {Glennon, Britta},
	year = {2024},
	pages = {907--930},
}

@article{nathan_wider_2014,
	title = {The wider economic impacts of high-skilled migrants: a survey of the literature for receiving countries},
	volume = {3},
	issn = {2193-9039},
	number = {1},
	journal = {IZA Journal of Migration},
	publisher = {Springer},
	author = {Nathan, Max},
	year = {2014},
	pages = {4},
}

@article{amuedo-dorantes_settling_2019,
	title = {Settling for academia?: {H}-{1B} visas and the career choices of international students in the {United} {States}},
	volume = {54},
	issn = {0022-166X},
	number = {2},
	journal = {Journal of Human Resources},
	publisher = {University of Wisconsin Press},
	author = {Amuedo-Dorantes, Catalina and Furtado, Delia},
	year = {2019},
	pages = {401--429},
}

@techreport{bernstein_contribution_2022,
	title = {The contribution of high-skilled immigrants to innovation in the {United} {States}},
	institution = {National Bureau of Economic Research},
	author = {Bernstein, Shai and Diamond, Rebecca and Jiranaphawiboon, Abhisit and McQuade, Timothy and Pousada, Beatriz},
	year = {2022},
}

@article{kerr_immigration_2013,
	title = {Immigration and employer transitions for {STEM} workers},
	volume = {103},
	issn = {0002-8282},
	number = {3},
	journal = {American Economic Review},
	publisher = {American Economic Association},
	author = {Kerr, Sari Pekkala and Kerr, William R},
	year = {2013},
	pages = {193--197},
}

@article{fonseca_unlocking_2024,
	title = {Unlocking {Mortgage} {Lock}-{In}: {Evidence} {From} a {Spatial} {Housing} {Ladder} {Model}},
	author = {Fonseca, Julia and Liu, Lu and Mabille, Pierre},
	year = {2024},
}

@article{gerardi_mortgage_2024,
	title = {Mortgage lock-in, lifecycle migration, and the welfare effects of housing market liquidity},
	publisher = {FRB Atlanta Working Paper},
	author = {Gerardi, Kristopher and Qian, Franklin and Zhang, David},
	year = {2024},
}

@article{aastveit_asymmetric_2022,
	title = {Asymmetric effects of monetary policy in regional housing markets},
	volume = {14},
	issn = {1945-7707},
	number = {4},
	journal = {American Economic Journal: Macroeconomics},
	publisher = {American Economic Association 2014 Broadway, Suite 305, Nashville, TN 37203-2425},
	author = {Aastveit, Knut Are and Anundsen, André K},
	year = {2022},
	pages = {499--529},
}

@article{howard_jue_2026,
	title = {{JUE} insight: {Moving} cost magnitudes in moving cost models},
	volume = {151},
	issn = {0094-1190},
	journal = {Journal of Urban Economics},
	publisher = {Elsevier},
	author = {Howard, Greg},
	year = {2026},
	pages = {103827},
}

@article{ioannides_housing_2025,
	title = {Housing and inequality},
	volume = {63},
	issn = {0022-0515},
	number = {3},
	journal = {Journal of Economic Literature},
	publisher = {American Economic Association 2014 Broadway, Suite 305, Nashville, TN 37203-2425},
	author = {Ioannides, Yannis M and Ngai, L Rachel},
	year = {2025},
	pages = {916--963},
}

@article{brown_locked_2020,
	title = {Locked in by leverage: {Job} search during the housing crisis},
	volume = {136},
	issn = {0304405X},
	shorttitle = {Locked in by leverage},
	url = {https://linkinghub.elsevier.com/retrieve/pii/S0304405X19302697},
	doi = {10.1016/j.jfineco.2019.11.001},
	language = {en},
	number = {3},
	urldate = {2026-01-07},
	journal = {Journal of Financial Economics},
	author = {Brown, Jennifer and Matsa, David A.},
	month = jun,
	year = {2020},
	pages = {623--648},
}

@article{howard_effects_2025,
	title = {The {Effects} of {Credit} {Access} on {Mobility} and {Neighborhood} {Choice}},
	author = {Howard, Greg and Liebersohn, Jack and Rodrigues, Flavio},
	year = {2025},
}

@article{fonseca_mortgage_2024,
	title = {Mortgage {Lock}‐{In}, {Mobility}, and {Labor} {Reallocation}},
	volume = {79},
	issn = {0022-1082, 1540-6261},
	url = {https://onlinelibrary.wiley.com/doi/10.1111/jofi.13398},
	doi = {10.1111/jofi.13398},
	language = {en},
	number = {6},
	urldate = {2026-01-06},
	journal = {The Journal of Finance},
	author = {Fonseca, Julia and Liu, Lu},
	month = dec,
	year = {2024},
	pages = {3729--3772},
}

@article{liebersohn_household_2025,
	title = {Household mobility and mortgage rate lock},
	volume = {164},
	issn = {0304405X},
	url = {https://linkinghub.elsevier.com/retrieve/pii/S0304405X2400196X},
	doi = {10.1016/j.jfineco.2024.103973},
	language = {en},
	urldate = {2026-01-06},
	journal = {Journal of Financial Economics},
	author = {Liebersohn, Jack and Rothstein, Jesse},
	month = feb,
	year = {2025},
	pages = {103973},
}

@article{dorn_essays_2009,
	title = {Essays on {Inequality}, {Spatial} {Interaction}, and the {Demand} for {Skills}},
	journal = {Dissertation, University of St. Gallen},
	author = {Dorn, David},
	year = {2009},
}
